\newif\ifArxivFormat
\ArxivFormattrue

\ifArxivFormat
\PassOptionsToPackage{pagebackref}{hyperref}
\documentclass{ioparxiv}
\else
\documentclass[anonymous]{iopjournal}
\fi

\usepackage[utf8]{inputenc}
\usepackage[T1]{fontenc}
\usepackage{lmodern}

\usepackage{amsmath, amssymb}
\usepackage{bbm}
\usepackage{graphicx}
\usepackage{booktabs}
\usepackage{capt-of}
\usepackage{siunitx}
\usepackage{multirow}

\ifArxivFormat
\usepackage[square,numbers]{natbib}
\else
\usepackage[round,authoryear]{natbib}
\makeatletter
\def\NAT@force@numbers{}
\makeatother
\fi

\usepackage{xcolor}
\definecolor{refblue}{RGB}{82,154,255}
\usepackage{url}
\usepackage[final]{microtype}
\usepackage{placeins}
\usepackage{wrapfig}
\usepackage{etoolbox}
\hypersetup{
  colorlinks=true,
  breaklinks=true,
  citecolor=refblue,
  linkcolor=red,
  urlcolor=refblue
}
\AtBeginDocument{%
  \hypersetup{
    citecolor=refblue,
    linkcolor=red,
    urlcolor=refblue
  }%
}
\ifArxivFormat
\providecommand*{\backref}[1]{}
\providecommand*{\backrefalt}[4]{}

\renewcommand*{\backref}[1]{}
\renewcommand*{\backrefalt}[4]{%
  \ifcase#1 %
  \or\space#2%
  \else\space#2%
  \fi
}

\fi

\AtBeginEnvironment{thebibliography}{%
  \small
  \setlength{\itemsep}{1pt}%
  \setlength{\parskip}{0pt}%
  \setlength{\parsep}{0pt}%
}

\newcommand{\FigPanel}[2]{Fig.~\hyperref[#1]{\ref*{#1}#2}}
\newcommand{\FigPanelRange}[3]{Fig.~\hyperref[#1]{\ref*{#1}#2--\ref*{#1}#3}}
\newcommand{\VideoRef}[1]{\hyperlink{vid:#1}{Video~#1}}

\begin{document}

\title{4D Vessel Reconstruction for Benchtop Thrombectomy Analysis}

\author{Ethan Nguyen$^{1,2}$\orcid{0009-0002-4774-1706}, Javier Carmona, Ph.D.$^3$\orcid{0000-0001-8574-2270}, Arisa Matsuzaki$^2$\orcid{0009-0005-7877-1459}, Naoki Kaneko, M.D., Ph.D.$^{4,*}$\orcid{0000-0002-3579-7908} and Katsushi Arisaka, Ph.D.$^{2,*}$\orcid{0000-0002-9585-4273}\linebreak}

\affil{$^1$UCLA Health}

\affil{$^2$UCLA Physics and Astronomy}

\affil{$^3$Chan Zuckerberg Biohub Network}

\affil{$^4$Ronald Reagan UCLA Medical Center}

\affil{$^*$Authors to whom any correspondence should be addressed.}

\email{ethunguy@ucla.edu, arisaka@physics.ucla.edu; nkaneko@mednet.ucla.edu}

\keywords{Mechanical thrombectomy, Gaussian Splatting, Multi-view 3D reconstruction, 4D reconstruction, Cerebrovascular phantom, Surface stress analysis, Synthetic validation, Biomechanical imaging}

\begin{abstract}

  \textbf{Introduction:} Mechanical thrombectomy can cause vessel deformation and procedure-related injury. Benchtop models are widely used for device testing, but time-resolved, full-field 3D vessel-motion measurements remain limited.

  \textbf{Methods:} We developed a nine-camera, low-cost multi-view workflow for benchtop thrombectomy in silicone middle cerebral artery phantoms (2160p, 20~fps). Multi-view videos were calibrated, segmented, and reconstructed with 4D Gaussian Splatting. Reconstructed point clouds were converted to fixed-connectivity edge graphs for region-of-interest (ROI) displacement tracking and a relative surface-based stress proxy. Stress-proxy values were derived from edge stretch using a Neo-Hookean mapping and reported as comparative surface metrics. A synthetic Blender pipeline with known deformation provided geometric and temporal validation.

  \textbf{Results:} In synthetic bulk translation, the stress proxy remained near zero for most edges (median \(\approx\) \SI{0}{MPa}; 90th percentile \SI{0.028}{MPa}), with sparse outliers. In synthetic pulling (1--\SI{5}{mm}), reconstruction showed close geometric and temporal agreement with ground truth, with symmetric Chamfer distance of 1.714--\SI{1.815}{mm} and precision of 0.964--0.972 at \(\tau=\SI{1}{\milli\meter}\). In preliminary benchtop comparative trials (one trial per condition), cervical aspiration catheter placement showed higher max-median ROI displacement and stress-proxy values than internal carotid artery terminus placement.

  \textbf{Conclusion:} The proposed protocol provides standardized, time-resolved surface kinematics and comparative relative displacement and stress proxy measurements for thrombectomy benchtop studies. The framework supports condition-to-condition comparisons and methods validation, while remaining distinct from absolute wall-stress estimation.\ifArxivFormat\ Implementation code and example data are available at \url{https://ethanuser.github.io/vessel4D}.\fi
\end{abstract}

\section{Introduction}
Mechanical thrombectomy is an effective treatment for acute ischemic stroke due to large-vessel occlusion, but procedure-related vessel injury and hemorrhagic complications remain clinically important. Recent work has highlighted post-procedural subarachnoid hemorrhage (SAH) as a meaningful complication associated with vessel overextension or displacement during retrieval, while multicenter data show that vessel perforation, although relatively uncommon, is associated with poor functional outcomes and higher mortality \citep{ishiguroLowRadialAxial2025,dmytriwIncidenceClinicalOutcomes2024}. These risks motivate experimental methods that can localize where deformation concentrates during retrieval maneuvers rather than relying only on aggregate procedural outcomes.

Benchtop and in-vitro thrombectomy platforms provide controlled and repeatable environments for device development, training, and mechanistic testing \citep{liuPreclinicalTestingPlatforms2021,johnsonReviewAdvancementsInvitro2022}. However, many current benchtop measurements emphasize recanalization, distal embolization, clot--device interaction, or force/friction metrics, rather than time-resolved, full-field 3D vessel-surface kinematics. Recent examples include studies of stent-retriever removal forces in tortuous models, friction-focused analyses, vessel deviation measured at a limited set of landmarks, and objective quantification of thrombus deformation during retrieval \citep{poulosInvestigationStentRetriever2024,nagargojeRoleFrictionForces2025,ishiguroLowRadialAxial2025,ernstMakingInvisibleVisible2026}. This leaves a gap between global or sparse measurements and spatially resolved characterization of vessel-surface deformation at anatomically meaningful regions during thrombectomy.

Existing optical and imaging approaches also pose practical constraints for this application. Multi-view 3D digital image correlation (3D-DIC) can provide full-field deformation measurements, but it depends on stable visible surface texture or applied speckle patterns, accurate multi-camera calibration, and adequate view coverage across time \citep{solavMultiDICOpenSourceToolbox2018,palancaUseDigitalImage2016,sanchezAssessmentFilmFreeWater2024}. More broadly, dynamic multi-view reconstruction requires explicit treatment of temporal coherence, since temporally unstable geometry is a recognized challenge in dynamic scene reconstruction \citep{mustafaTemporallyCoherentGeneral2021,wu4DGaussianSplatting2024}. Clinical dynamic CT methods such as 4D-CTA address a different imaging niche in neurovascular imaging and are not the focus of the present benchtop optical workflow \citep{kortman4DCTANeurovascularDisease2015}.

This study presents a multi-view dynamic reconstruction protocol for silicone middle cerebral artery (MCA) thrombectomy phantoms, combining multi-view camera acquisition, segmentation, 4D Gaussian Splatting reconstruction, fixed-connectivity edge-graph construction, region-of-interest (ROI) displacement tracking, and a relative surface-based stress proxy \citep{wu4DGaussianSplatting2024}. The stress proxy is used as a comparative surface metric and is not interpreted as absolute wall stress.

The study aims are: (1) to define a standardized end-to-end acquisition-to-analysis protocol for dynamic benchtop thrombectomy imaging; (2) to formalize ROI-level displacement and relative stress-proxy metrics on a fixed deforming edge graph; (3) to validate geometric and temporal behavior in synthetic data with known motion; and (4) to demonstrate comparative benchtop application in two aspiration catheter (AC) placement conditions (cervical internal carotid artery (ICA) vs ICA terminus).

\section{Materials and Methods}

\subsection{Pipeline overview}
\begin{figure}[!htbp]
  \centering
  \includegraphics[width=0.8\linewidth]{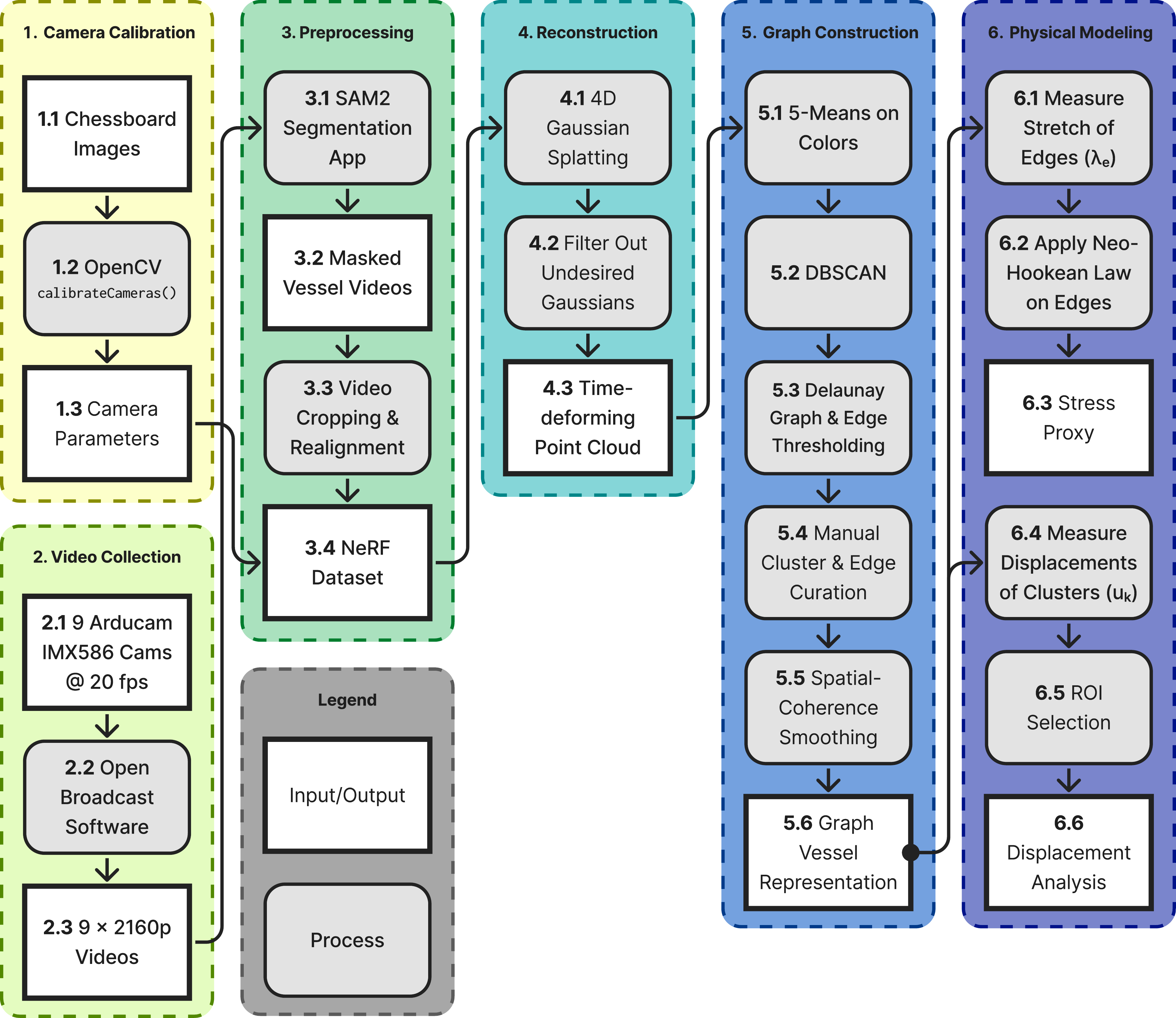}
  \caption{Overview of the experimental and computational pipeline. Multi-camera chessboard calibration yields camera parameters for nine-view vessel recordings. After SAM2-based segmentation, cropping, and dataset preparation, the deforming vessel is reconstructed with 4D Gaussian Splatting and filtered to obtain a time-varying point cloud. The point cloud is converted into a fixed-connectivity edge graph using clustering, DBSCAN, Delaunay graph pruning, and one-time manual curation. Downstream outputs are region-of-interest displacement magnitude (mm) and edge-based stress-proxy magnitude $|\sigma_e|$ (MPa). Square boxes denote inputs/outputs and rounded gray boxes denote processing steps.}
  \label{fig:pipeline}
\end{figure}

The workflow comprises acquisition, calibration, segmentation, dynamic reconstruction, fixed-connectivity edge-graph generation, ROI metric extraction, and synthetic/benchtop evaluation (Fig.~\ref{fig:pipeline}). Representative qualitative outputs across these stages are shown in Fig.~\ref{fig:stages}. The benchtop phantom was a patient-specific silicone ICA/MCA model prepared as described previously~\citep{kanekoManufacturePatientspecificVascular2016}. Synthetic validation and benchtop experiments used a variant with an additional loop connecting the M2 branches, whereas Fig.~\ref{fig:stages} uses a different ICA/MCA silicone model without that loop for illustration only. Typical per-experiment end-to-end processing from multi-view video feeds to displacement/stress-proxy graph generation was approximately 20~minutes for automated reconstruction plus 10~minutes for user-guided preprocessing/curation.

\subsubsection{Benchtop demonstration scope}
To clarify interpretation of downstream outputs, benchtop data were used as a methods demonstration, and one representative trial was analyzed per AC placement condition. A Trevo NXT stent retriever was deployed from the inferior M2 branch to the M1/M2 bifurcation together with a Toro 88 aspiration catheter; no clot analog was present. Analysis was restricted to the retraction interval after pullback began. We defined \(t=0\) as linear-actuator pullback onset. Displayed peak-deformation frames were chosen by maximizing whole-vessel median displacement and whole-vessel median stress-proxy values; these maxima coincided in the shown experiments. The same processing and metric definitions were then applied across both conditions.

\subsubsection{Imaging hardware and acquisition}
A low-cost multi-view rig comprising nine Arducam IMX586 cameras was used to record raw 2160p video at 20~fps. Cameras were mounted on a dodecahedron-like frame constructed from PVC pipes, 3D-printed connecting joints, and custom 3D-printed camera mounts, with five cameras on the lower ring and four on the upper ring. The cameras were manually aimed toward a common central target region and fixed after calibration. Preliminary synthetic-data testing indicated that nine views were sufficient for 3D recovery, although additional views would be expected to further improve reconstruction quality. The imaging hardware (PVC structure, printed parts, mounts, fasteners, cameras, and lighting) cost less than \$1500 in total.

For the reported configuration, camera-to-target distances ranged from approximately 230 to 271~mm (mean \(\approx\) 247~mm). The M1 mid segment was positioned near the center of the shared field of view. Views were acquired through a tiled OBS composite workflow and later cropped to \(1200\times1200\) pixels, with crop centers aligned to a common chessboard origin to simplify downstream 4DGS processing. Although inter-camera skew was not independently hardware-validated, no visually apparent offset was observed across shared motion events, and the analyzed deformation evolved over seconds rather than impulsive frame-to-frame motion; residual mismatch was therefore expected to be small relative to the motion timescale studied here. The usable imaged volume after clipping was approximately \(70\times90\times60\)~mm (width \(\times\) height \(\times\) depth), although the effective working volume could be adjusted depending on camera focus and the region of interest. Approximately 20--30~s were recorded per experiment, then temporally downsampled so that roughly 1000 total frames were used for 4DGS processing. The rig layout and representative multi-view raw frames are shown in \FigPanelRange{fig:stages}{A}{B}. Representative raw benchtop footage is provided in \VideoRef{S1}.

\subsubsection{Illumination and surface feature preparation}
Red fluorescent microspheres (approximately \SI{300}{\micro\meter}) were applied to the vessel exterior. UV illumination (approximately \SI{400}{nm}) was used to improve feature contrast on the silicone surface, as illustrated by the representative raw frames in \FigPanel{fig:stages}{B}. This preparation improved visible surface texture before segmentation.

\subsubsection{Multi-camera calibration and metric scaling}
Extrinsics were estimated using planar chessboard calibration~\citep{zhangFlexibleNewTechnique2000} implemented with OpenCV \texttt{calibrateCamera}~\citep{opencv_library} from multiple board orientations using a \(6\times8\) board, with the camera calibration process shown in \FigPanel{fig:stages}{A}. Additional side-by-side nine-view calibration panels for real and synthetic setups are provided in Supplementary Fig.~\ref{fig:supp_calibration_pattern}. The mean reprojection error across cameras was approximately 0.134 pixels (range 0.112--0.164 pixels). Metric scaling was obtained by matching reconstructed chessboard geometry to measured physical board dimensions. These calibrated parameters were used directly in 4DGS reconstruction.

\subsubsection{Deformation apparatus and pullback protocol}
A linear actuator imposed controlled pullback at \SI{4}{\milli\meter\per\second}. The vessel trunk/base was fixed relative to the camera rig to reduce global drift during acquisition. The analyzed interval was the retraction maneuver after pullback onset. This same temporal reference was used throughout downstream tracking and metric extraction.

\subsubsection{Video segmentation and background suppression}
Each camera stream was segmented independently using SAM2 video prediction (\texttt{sam2.1\_hiera\_large.pt})~\citep{raviSAM2Segment2024}. For each view, prompts were placed once on the first frame and propagated through the sequence; masks were reinitialized only when flicker or incorrect vessel coverage appeared. The same segmentation protocol was used for all sequences. Non-vessel pixels were set to black, and no additional mask post-processing beyond cropping/alignment was applied (representative output in \FigPanel{fig:stages}{C}). The resulting masked frames were then passed to dynamic reconstruction.

\subsubsection{Dynamic reconstruction}
Calibrated multi-view frames were reconstructed with 4D Gaussian Splatting (4DGS) using the implementation of Wu et al.~\citep{wu4DGaussianSplatting2024}. One model was trained per sequence for 20,000 iterations on an NVIDIA RTX 4070 Ti GPU, yielding the representative dynamic point cloud shown in \FigPanel{fig:stages}{D}. Exports contained Gaussian primitive centers and colors over time; primitives with small radii, low opacity, or no color were filtered out before downstream graph construction. These filtered primitive trajectories were then converted to a fixed graph representation.

\subsubsection{Point-cloud post-processing and fixed edge-graph construction}
To support consistent temporal metric extraction, reconstructed primitives were converted to a fixed vertex set and static edge-graph connectivity. For each sequence, frames are indexed by \(t=0,\dots,T-1\), and the resulting clustered vertices and graph structure are illustrated in \FigPanel{fig:stages}{E}.

Initial clustering used KMeans in RGB (\(K_c=5\); \citep{macqueenMethodsClassificationAnalysis1967}) followed by DBSCAN (\citep{esterDensitybasedAlgorithmDiscovering1996}), implemented in scikit-learn~\citep{pedregosaScikitlearnMachineLearning2012}, within each color group, with \(\varepsilon_s=\SI{0.7}{mm}\) (eps) and \(m_s=3\) (minPts). This two-stage strategy was used first to prevent points with different colors from being grouped together, and then to separate spatially distinct local components within each color group for downstream tracking. To formalize this procedure, let \(\mathbf{p}_n(t)\in\mathbb{R}^3\) denote the XYZ position of primitive \(n\) at frame \(t\), and let \(g_n\in\{1,\dots,K_c\}\) denote its KMeans color-group label at \(t=0\). Formally, for color group \(c\):
\begin{equation}
  \{\mathbf{p}_n(0): g_n=c\}
  \xrightarrow{\ \mathrm{DBSCAN}(\varepsilon_s, m_s)\ }
  \{\mathcal{C}_{c,r}\}_{r=1}^{R_c},
  \label{eq:dbscan_components_main}
\end{equation}
where \(\mathcal{C}_{c,r}\) are DBSCAN-connected spatial components (clusters) within color group \(c\). Cluster memberships \(\ell_n\in\{1,\dots,K\}\) from \(t=0\) were held fixed, where \(K\) is the number of retained clusters, and vertex positions were computed by centroid averaging:
\begin{equation}
  \mathbf{x}_k(t)=\frac{1}{|\mathcal{I}_k|}\sum_{n\in\mathcal{I}_k}\mathbf{p}_n(t),
  \qquad
  \mathcal{I}_k=\{n:\ell_n=k\},
  \label{eq:cluster_centroid}
\end{equation}
with displacement relative to baseline:
\begin{equation}
  \mathbf{u}_k(t)=\mathbf{x}_k(t)-\mathbf{x}_k(0).
  \label{eq:vertex_displacement}
\end{equation}
If \(\mathcal{I}_k=\emptyset\) for a frame, the previous vertex position was retained.

Static connectivity was generated on \(\{\mathbf{x}_k(0)\}_{k=1}^K\) using a 3D Delaunay neighborhood graph (SciPy spatial routines~\citep{virtanenSciPy10Fundamental2020}) and edge-length pruning. Candidate edges \((i,j)\) were retained when:
\begin{equation}
  d_{ij} \le \mu_d + \gamma\,\sigma_d,
  \qquad d_{ij}=\|\mathbf{x}_i(0)-\mathbf{x}_j(0)\|_2,
  \label{eq:edge_pruning_main}
\end{equation}
where \(\mu_d\) and \(\sigma_d\) are the candidate edge-length mean and standard deviation; \(\gamma=0.25\) in reported analyses. Let \(\mathcal{E}\) denote the retained edge set of the pruned Delaunay graph. Neighborhoods \(\mathcal{N}(k)\) for spatial coherence were defined by adjacency in this pruned graph at \(t=0\). This fixed graph representation was then used for curation, smoothing, and ROI aggregation.

\subsubsection{Manual curation and topology locking}
To reduce obvious topological artifacts before temporal analysis, manual curation was performed once on the initial frame using a custom PyVista-based 3D editor to remove extraneous clusters, standardize vessel crop length, and remove implausible edges (e.g., edges spanning nearby trunks). The curated cluster set and edge topology were then reused unchanged for all subsequent frames.

\subsubsection{Spatial coherence filtering}
To suppress isolated local tearing in the tracked displacement field before ROI-level aggregation, we applied a mild spatial regularization on the fixed cluster-adjacency graph defined above, using the curated saved edge set when available and otherwise the pruned Delaunay graph. This update can be viewed as relaxed graph/Laplacian smoothing (low-pass fairing), while the residual-dependent weights limit smoothing across locally inconsistent motions, in the spirit of anisotropic diffusion and robust piecewise-smooth motion regularization \citep{taubinSignalProcessingApproach1995,peronaScalespaceEdgeDetection1990,blackRobustAnisotropicDiffusion1998,blackRobustEstimationMultiple1996}. Standard reported analyses used one iteration with \(\alpha=0.1\), robust weighting enabled, \(\kappa=2.5\), and displacement reference \(\mathbf{x}_k(0)\) (``\(t_0\)'' reference mode in code), although the number of smoothing iterations is configurable. For each frame, with residuals and neighbor averages recomputed at each iteration,
\begin{equation}
  \mathbf{u}_k \leftarrow (1-\alpha)\mathbf{u}_k
  + \alpha\,
  \frac{\sum_{j\in\mathcal{N}(k)} w_{kj}\mathbf{u}_j}{\sum_{j\in\mathcal{N}(k)} w_{kj}+\varepsilon},
  \label{eq:coherence_update_main}
\end{equation}
with residuals \(r_{kj}=\|\mathbf{u}_k-\mathbf{u}_j\|_2\) and robust weights
\begin{equation}
  w_{kj}=
  \begin{cases}
    1, & r_{kj}\le\tau,\\
    \tau/(r_{kj}+\varepsilon), & r_{kj}>\tau,
  \end{cases}
  \qquad \tau=\kappa s,
  \label{eq:coherence_weight_main}
\end{equation}
where the weight function corresponds to a Huber-like IRLS weighting, so neighbors with similar displacements are averaged more strongly whereas neighbors with large local disagreements are downweighted rather than fully trusted \citep{huberRobustEstimationLocation1964}. The robust scale is computed once per frame over edges using a median-absolute-deviation (MAD) estimate,
\begin{equation}
  s=\operatorname{median}_{(i,j)\in\mathcal{E}}\left|r_{ij}-\operatorname{median}_{(a,b)\in\mathcal{E}} r_{ab}\right|+\varepsilon,
  \label{eq:coherence_scale_main}
\end{equation}
which provides an adaptive threshold that is less sensitive to outlying residuals than variance-based scaling \citep{rousseeuwAlternativesMedianAbsolute1993}. Here \(\varepsilon=10^{-12}\) is a numerical stabilizer used in code for both the MAD scale and denominator safeguards. Clusters with no points in a frame retained their previous positions and were not themselves updated by the smoothing step. The filtered displacements were then used for ROI displacement and stress-proxy computation. To assess the effect of this step, we also repeated the synthetic validation analysis without spatial coherence filtering; the corresponding ablation results are provided in Supplementary Fig.~\ref{fig:supp_agreement_no_scf}, Supplementary Fig.~\ref{fig:supp_scf_compare_qual}, and Supplementary Table~\ref{tab:scf_ablation_compare}.

\subsubsection{ROI displacement and stress-proxy metrics}\label{sec:ROI_metrics}
\begin{figure}[!htbp]
  \centering
  \includegraphics[width=0.25\linewidth]{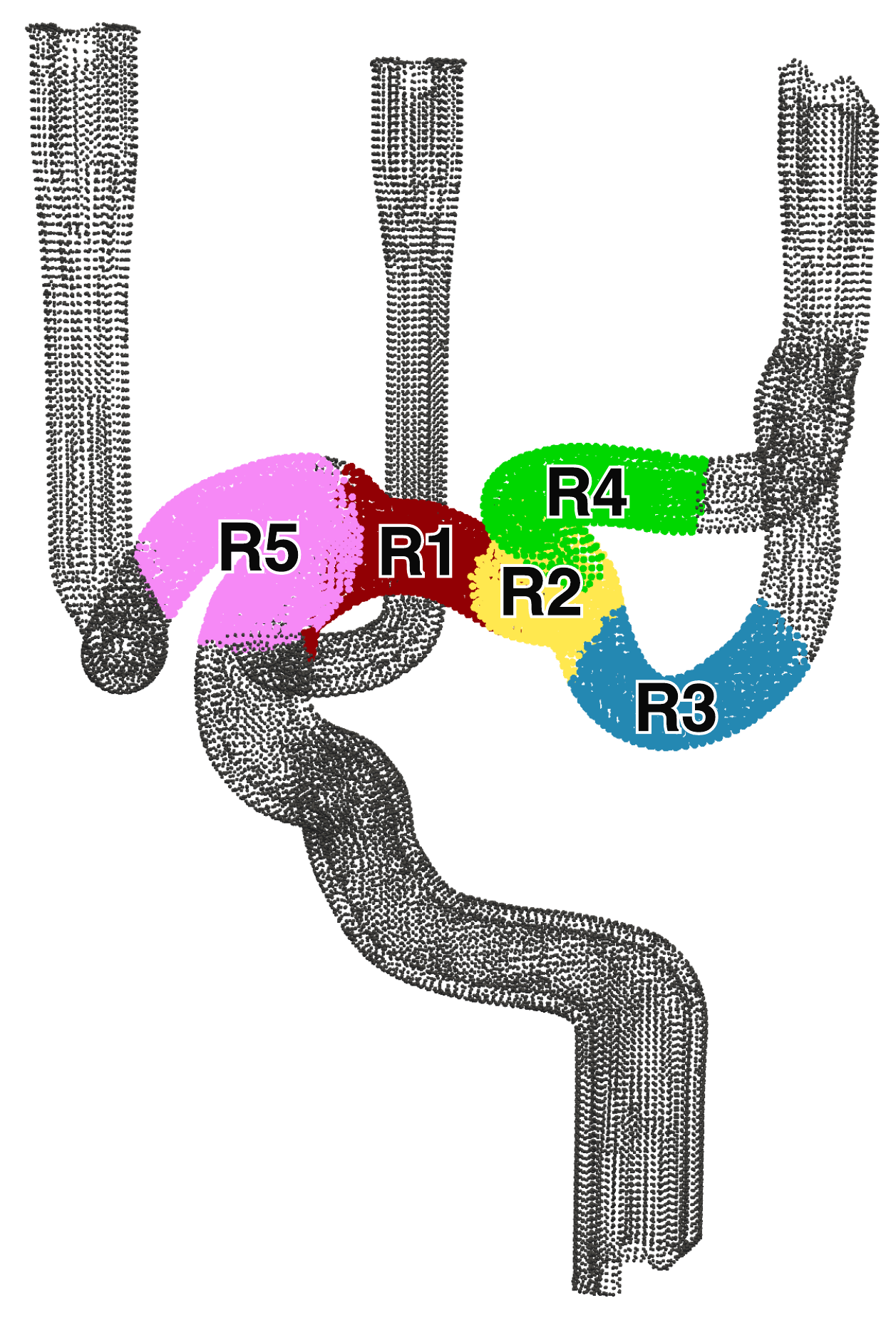}
  \caption{Regions of the ICA/MCA surface used for regional aggregation and analysis. R1: M1 mid segment; R2: M1/M2 bifurcation; R3: proximal inferior M2; R4: proximal superior M2; R5: ICA terminus/M1 origin.}
  \label{fig:regions}
\end{figure}

At this stage, graph trajectories were summarized into ROI-level outputs for comparison. ROIs were defined on anatomical landmarks (Fig.~\ref{fig:regions}); each ROI denotes a selected set of vertex indices on the fixed graph. For benchtop data, ROIs were manually selected by the researcher on the initial fixed graph and reused over time, with placement matched as closely as possible across conditions. For synthetic data, ROIs were defined as sphere-based neighborhoods around fixed centers and reused for both ground truth (GT) and reconstruction because initial synthetic frames were spatially aligned. Per-frame ROI displacement was the vertex-wise median magnitude:
\begin{equation}
  d^{\mathrm{med}}_{\mathrm{ROI}}(t)=\operatorname{median}_{k\in\mathrm{ROI}}\|\mathbf{u}_k(t)\|_2.
  \label{eq:roi_displacement}
\end{equation}
Condition-level displacement summary used the temporal maximum of this median:
\begin{equation}
  d^{\max}_{\mathrm{ROI}}=\max_t d^{\mathrm{med}}_{\mathrm{ROI}}(t).
  \label{eq:roi_displacement_max}
\end{equation}

\label{sec:stress_proxy}
To map edge deformation into a comparative stress-like quantity, for each edge \(e=(i,j)\in\mathcal{E}\) in the fixed connectivity graph let \(\ell_e(t)=\|\mathbf{x}_i(t)-\mathbf{x}_j(t)\|_2\) denote the edge length at frame \(t\). Stretch was then defined by:
\begin{equation}
  \lambda_e(t) = \frac{\ell_e(t)}{\ell_e(0)},
  \label{eq:edge_stretch}
\end{equation}
and mapped to a Neo-Hookean uniaxial stress-like quantity. For incompressible isotropic hyperelasticity with traction-free lateral contraction, the uniaxial Cauchy stress reduces to the Neo-Hookean form~\citep{holzapfelNonlinearSolidMechanics2000}:
\begin{equation}
  \sigma_e = \mu\left(\lambda_e^2 - \lambda_e^{-1}\right),
  \label{eq:neohookean_uniaxial}
\end{equation}
with \(\mu\) derived from silicone Young's modulus \(E\). For near-incompressible material, \(\mu = E/[2(1+\nu)] \approx E/3\) at \(\nu\approx0.5\). Using \(E=\SI{1.15}{\mega\pascal}\) for silicone~\citep{chuehNeurovascularModelingSmallBatch2009} gives \(\mu\approx\SI{0.383}{\mega\pascal}\).

\begin{figure}[!htbp]
  \centering
  \includegraphics[width=\linewidth]{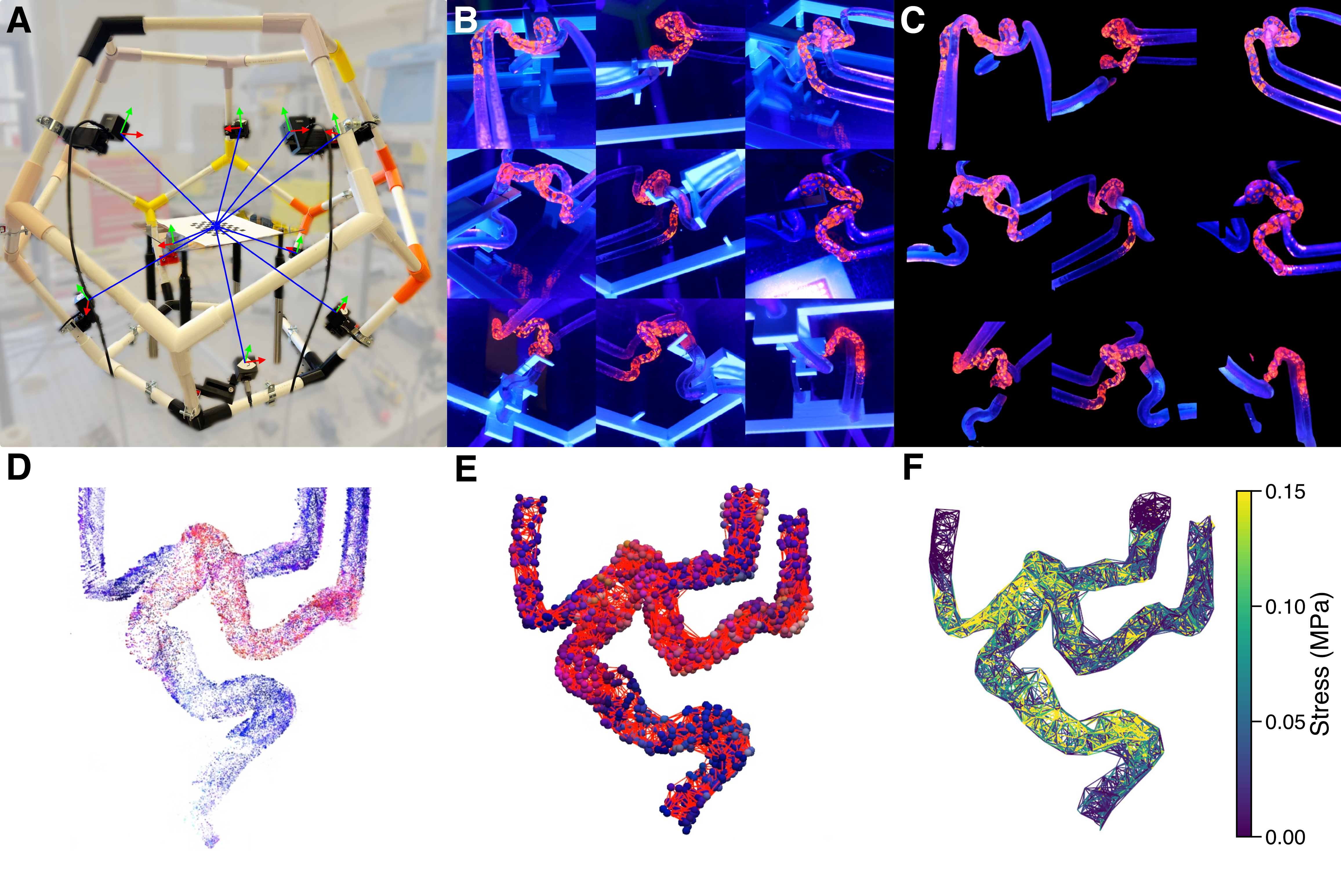}
  \caption{Experimental setup and major stages of the reconstruction and analysis pipeline.
    (A) Nine-camera acquisition rig used for multi-view imaging, shown during chessboard-based calibration; long arrows (shown in blue) indicate the cameras' viewing directions toward the working volume.
    (B) Example multi-view raw frames acquired during benchtop thrombectomy imaging under UV illumination, showing the fluorescent-beaded silicone vessel phantom.
    (C) Corresponding segmented vessel frames after preprocessing with SAM2, used to isolate vessel for downstream reconstruction.
    (D) Example time-varying 3D point-cloud reconstruction obtained from 4D Gaussian Splatting.
    (E) Fixed set of tracked surface vertices obtained from point-cloud clustering and used for fixed edge-graph construction and temporal tracking.
  (F) Edge-based stress-proxy magnitude $|\sigma_e|$ (MPa) on the resulting edge graph at the representative timepoint.}
  \label{fig:stages}
\end{figure}

Per-frame ROI stress summary used the median of \(|\sigma_e|\) over edges with both endpoints in the ROI, with condition-level max-median reporting analogous to Eq.~\ref{eq:roi_displacement_max}. Representative stress-proxy output is shown in \FigPanel{fig:stages}{F}. The stress proxy is interpreted as a relative, surface-based comparative metric and not as absolute wall stress. These ROI summaries were then carried into synthetic and benchtop comparisons.

\subsection{Synthetic validation design and metric definitions}\label{sec:synthetic_validation_metrics}
Synthetic experiments were used to check geometric and temporal behavior against known motion before interpreting benchtop comparisons.
\begin{figure}[!htbp]
  \centering
  \includegraphics[width=0.8\linewidth]{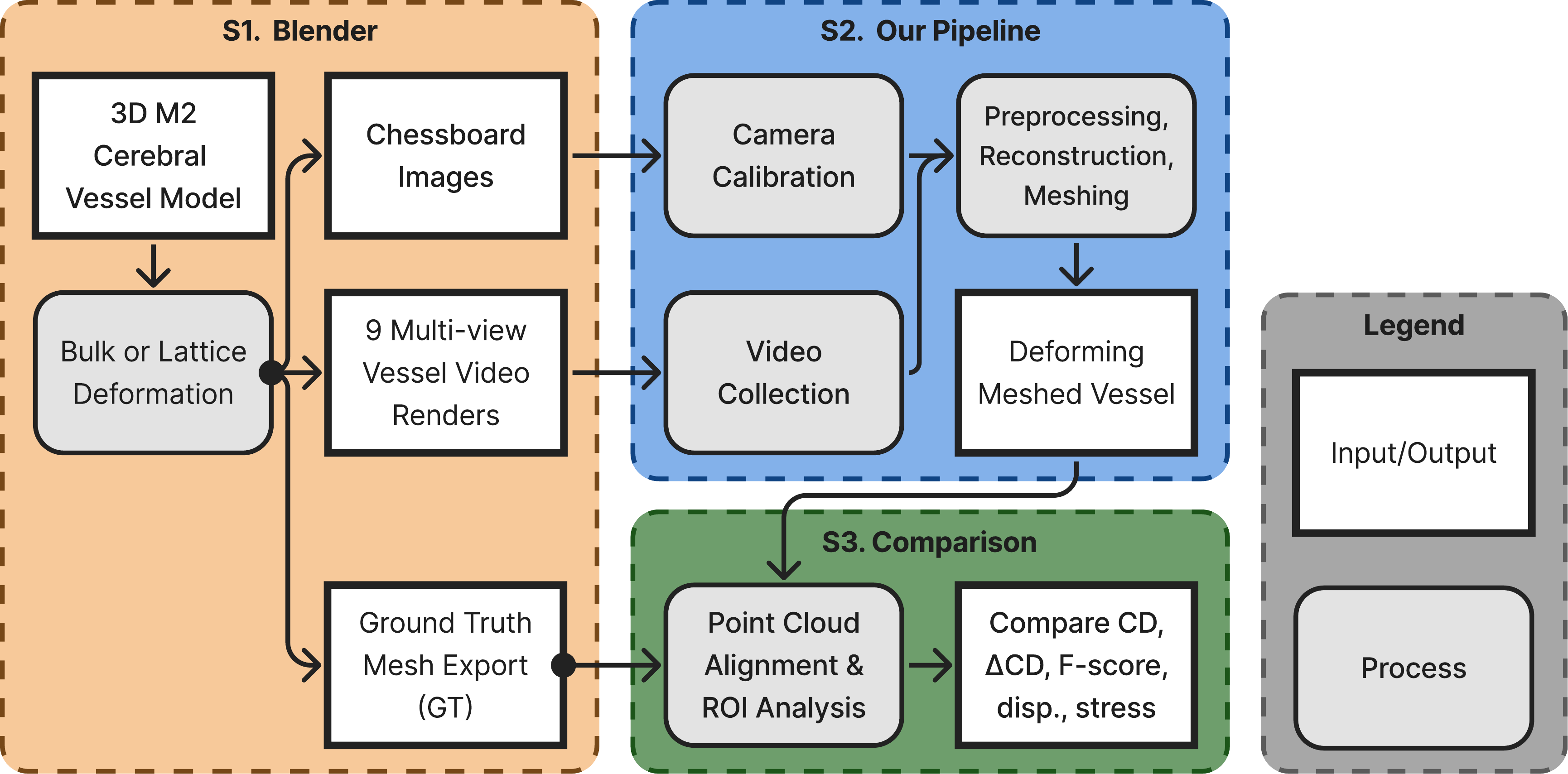}
  \caption{Synthetic validation workflow (schematic; no single frame/timepoint metric shown). Ground-truth vessel geometries were deformed in Blender using controlled bulk-translation and localized-pulling conditions, rendered with the same camera geometry as the physical setup, and passed through the same camera geometry and downstream reconstruction pipeline, with segmentation bypassed because the renders use transparent backgrounds. The resulting reconstructed point clouds and fixed edge graphs were compared with ground truth using Chamfer distance, temporal consistency metrics, and deformation/stress-proxy agreement analyses.}
  \label{fig:synth_pipeline}
\end{figure}

To enable quantitative validation with ground truth, a known vessel surface mesh was rendered in Blender with matched camera geometry (Fig.~\ref{fig:synth_pipeline}) under two conditions: bulk translation and localized pulling near the M1/M2 bifurcation (1--\SI{5}{mm} nominal pulls). Because these Blender renders used transparent backgrounds, this validation isolates the reconstruction-and-analysis stages and does not test the SAM~2-based background-filtering/segmentation step required for benchtop videos. The synthetic vessel appearance also only approximates the real phantom's optical surface characteristics, including vessel-surface contrast and bead size/visibility. Metrics used for comparison followed established dynamic-surface evaluation definitions and protocol~\citep{gongSWiT4DSlidingWindowTransformer2025}. GT stress-proxy maps were computed from the Blender-exported ground-truth edge graph using the same stress-proxy definition (Eqs.~\ref{eq:edge_stretch}--\ref{eq:neohookean_uniaxial}).

Symmetric Chamfer distance (CD) was defined as:
\begin{equation}
  \mathrm{CD}(P,G)=\frac{1}{\lvert P\rvert}\sum_{p\in P}\min_{g\in G}\lVert p-g\rVert_2
  +\frac{1}{\lvert G\rvert}\sum_{g\in G}\min_{p\in P}\lVert g-p\rVert_2,
  \label{eq:chamfer}
\end{equation}
where \(P\) and \(G\) are reconstructed and ground-truth point sets. We also report \(\mathrm{CD}_{\mathrm{norm}}=\mathrm{CD}/\tilde{d}_{\mathrm{GT}}\), where \(\tilde{d}_{\mathrm{GT}}\) is the median nearest-neighbor GT point spacing.

Non-temporal metrics (CD, \(\mathrm{CD}_{\mathrm{norm}}\), precision, recall, F-score) were evaluated at maximum deformation (last frame, \(t=T-1\)). Temporal consistency used frame-to-frame CD disagreement, following the same protocol~\citep{gongSWiT4DSlidingWindowTransformer2025}. For last-frame non-temporal metrics, directed distances were:
\begin{equation}
  d_{P\to G}(p)=\min_{g\in G_{T-1}} \lVert p-g\rVert_2,
  \label{eq:directed_dist_p2g}
\end{equation}
\begin{equation}
  d_{G\to P}(g)=\min_{p\in P_{T-1}} \lVert g-p\rVert_2,
  \label{eq:directed_dist_g2p}
\end{equation}
with temporal terms for \(t=0,\dots,T-2\):
\begin{equation}
  \mathrm{CD}^{\mathrm{pred}}_{t}=\mathrm{CD}(P_t,P_{t+1}),\qquad
  \mathrm{CD}^{\mathrm{gt}}_{t}=\mathrm{CD}(G_t,G_{t+1}),
  \label{eq:temporal_cd_defs_main}
\end{equation}
and temporal disagreement:
\begin{equation}
  \Delta \mathrm{CD} = \frac{1}{T-1}\sum_{t=0}^{T-2}\left\lvert
  \mathrm{CD}^{\mathrm{pred}}_{t}-\mathrm{CD}^{\mathrm{gt}}_{t}\right\rvert,
  \label{eq:delta_cd_main}
\end{equation}
reported also as \(\Delta\mathrm{CD}_{\mathrm{rel}}\) after normalization by the median ground-truth frame-to-frame CD.

At tolerance \(\tau=\SI{1}{mm}\) by default (evaluated at \(t=T-1\)),
\begin{align}
  \mathrm{Precision} &= \frac{1}{\lvert P_{T-1}\rvert}\sum_{p\in P_{T-1}} \mathbbm{1}\!\left[d_{P\to G}(p) < \tau\right], \\
  \mathrm{Recall} &= \frac{1}{\lvert G_{T-1}\rvert}\sum_{g\in G_{T-1}} \mathbbm{1}\!\left[d_{G\to P}(g) < \tau\right],
  \label{eq:pr_main}
\end{align}
and
\begin{equation}
  \mathrm{F\text{-}score} =
  \begin{cases}
    \dfrac{2\,\mathrm{Precision}\,\mathrm{Recall}}{\mathrm{Precision}+\mathrm{Recall}}, & \text{if } \mathrm{Precision}+\mathrm{Recall} > 0, \\
    0, & \text{otherwise.}
  \end{cases}
  \label{eq:fscore_main}
\end{equation}

Displacement error was defined as \(d^{\max}_{\mathrm{ROI,ours}}-d^{\max}_{\mathrm{ROI,gt}}\), with percent error computed relative to \(d^{\max}_{\mathrm{ROI,gt}}\). Stress bias was defined analogously as the Ours\(-\)GT difference in max-median ROI stress proxy. Regression and Bland--Altman analyses~\citep{blandStatisticalMethodsAssessing1986} used 25 paired points (5 ROIs \(\times\) 5 pull magnitudes). Together, these metrics define the synthetic benchmarks reported in Results.

\FloatBarrier

\section{Results}
Results are presented in the same sequence as the workflow: synthetic control, synthetic pulling validation, and benchtop comparison.

\subsection*{Synthetic rigid-motion rejection}
\begin{figure}[!htbp]
  \centering
  \includegraphics[width=0.5\linewidth]{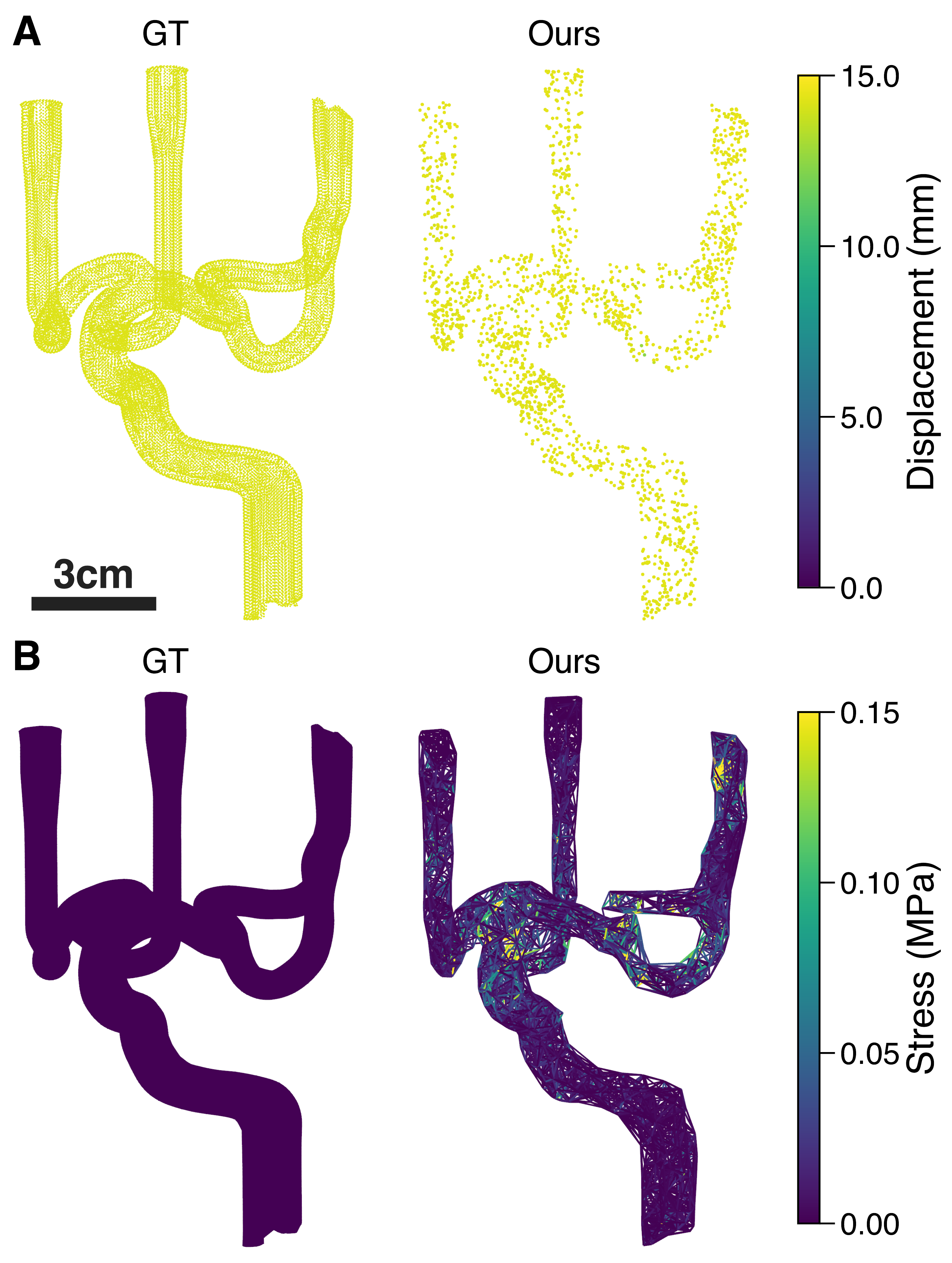}
  \caption{Rigid-body translation control (synthetic vessel), evaluated at the final frame of the bulk-translation sequence. (A) Displacement magnitude (mm) for ground truth (GT) and reconstruction (Ours). (B) Edge-based stress-proxy magnitude $|\sigma_e|$ (MPa); GT remains near zero under rigid motion, and Ours is near zero for most edges with sparse outliers.}
  \label{fig:synth_translation}
\end{figure}

The rigid-translation control yielded near-zero stress-proxy values for most edges: median \(\approx\) \SI{0}{MPa}, 90th percentile \SI{0.028}{MPa}, and a sparse outlier tail with maximum \SI{1.898}{MPa} (Fig.~\ref{fig:synth_translation}). Regional displacement error was \SI{0.091}{mm} (0.64\%), \(\Delta\mathrm{CD}_{\text{rel}}\) was 0.085, and stress bias was \SI{0.018}{MPa} (Table~\ref{tab:synthetic_validation}). Raw/displacement/stress videos are provided in \VideoRef{S6}, \VideoRef{S7}, and \VideoRef{S8}. With this control behavior established, we next evaluated localized pulling.

\subsection*{Synthetic deformation fidelity}
\begin{figure}[!htbp]
  \centering
  \includegraphics[width=\linewidth]{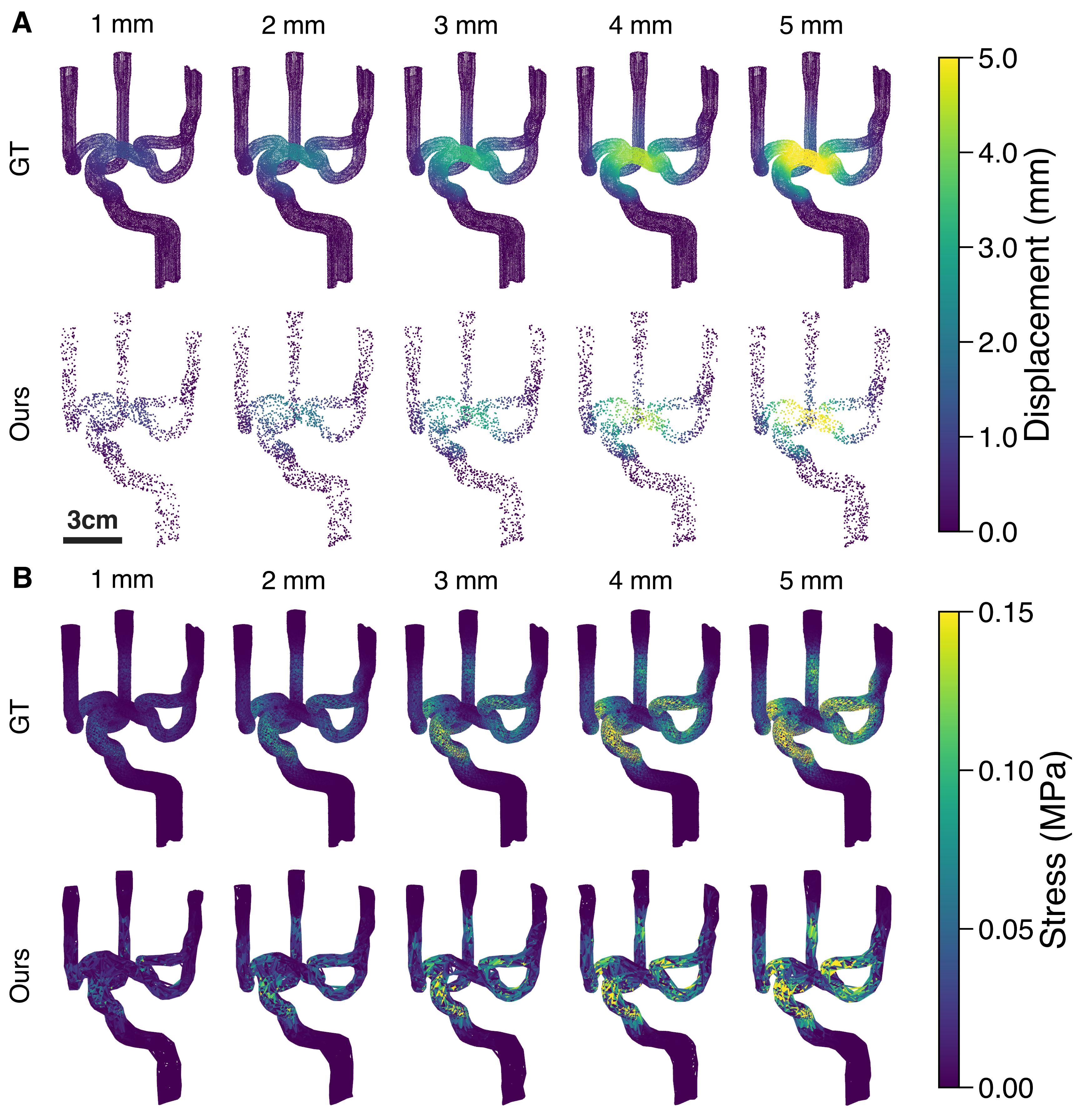}
  \caption{Synthetic pulling deformation (1--\SI{5}{mm}), evaluated at each condition's final frame. Rows compare ground truth (GT) and reconstruction (Ours). (A) Displacement magnitude (mm). (B) Edge-based stress-proxy magnitude $|\sigma_e|$ (MPa).}
  \label{fig:synth_pulling}
\end{figure}

Under localized pulling, reconstruction preserved condition-dependent deformation patterns with strong task-relevant agreement. Across 1--\SI{5}{mm} pulling conditions, symmetric CD ranged from 1.714 to \SI{1.815}{mm}, \(\mathrm{CD}_{\text{norm}}\) from 2.872 to 3.041, and \(\Delta\mathrm{CD}_{\text{rel}}\) from 0.100 to 0.139 (Table~\ref{tab:synthetic_validation} and Fig.~\ref{fig:synth_heatmaps}; expanded directed-distance and overlap metrics are provided in Supplementary Table~\ref{tab:synthetic_validation_supp}). Precision at \(\tau=\SI{1}{\milli\meter}\) ranged from 0.964 to 0.972. Regional displacement error ranged from \SI{0.047}{mm} to \SI{0.190}{mm}, and stress bias ranged from \SI{0.004}{MPa} to \SI{0.007}{MPa} (Fig.~\ref{fig:synth_pulling}).

\begin{figure}[!htbp]
  \centering
  \includegraphics[width=0.8\linewidth]{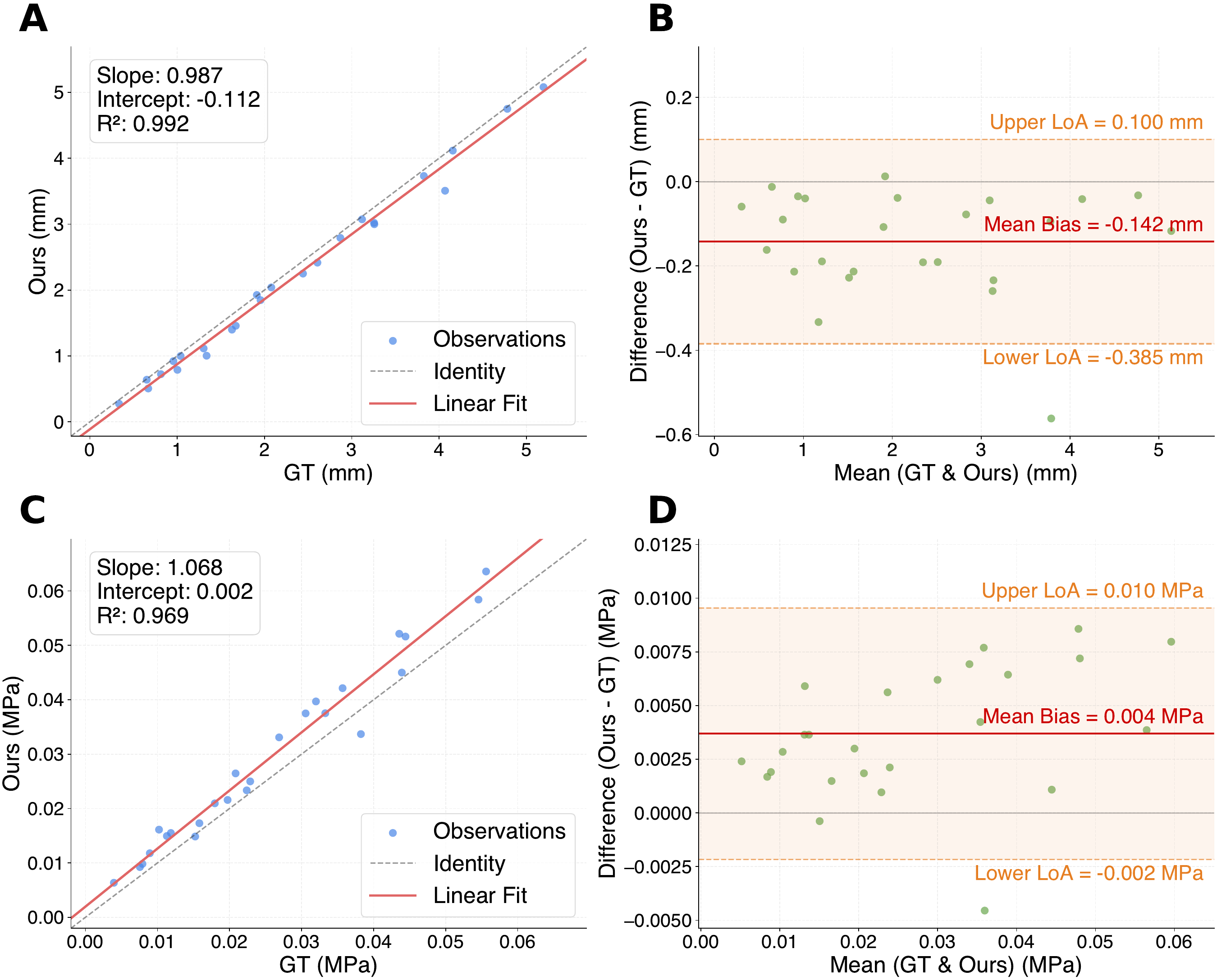}
  \caption{Agreement analyses for max-median ROI metrics between reconstruction and ground truth (condition-level summaries, not single-frame values): correlation and Bland--Altman for displacement magnitude (A--B, mm) and relative surface-based stress-proxy magnitude (C--D, MPa).}
  \label{fig:agreement}
\end{figure}

Agreement analysis of max-median ROI summaries showed displacement slope 0.987, intercept \(-\SI{0.112}{mm}\), and \(R^2=0.992\), with Bland--Altman bias \(-\SI{0.142}{mm}\) and limits \(-\SI{0.385}{mm}\) to \SI{0.100}{mm} (Fig.~\ref{fig:agreement}). Stress-proxy agreement showed slope 1.068, intercept \SI{0.002}{MPa}, and \(R^2=0.969\), with bias \SI{0.004}{MPa} and limits \(-\SI{0.002}{MPa}\) to \SI{0.010}{MPa}. Without spatial coherence filtering, displacement agreement remained similar, whereas stress-proxy agreement became less favorable, with larger positive bias and wider limits of agreement (Supplementary Fig.~\ref{fig:supp_agreement_no_scf}, Supplementary Fig.~\ref{fig:supp_scf_compare_qual}, and Supplementary Table~\ref{tab:scf_ablation_compare}). Full ROI-wise condition summaries are provided in Supplementary Table~\ref{tab:regional_summary}. Raw/displacement/stress videos are provided in \VideoRef{S9}, \VideoRef{S10}, and \VideoRef{S11}. These synthetic findings motivate the same ROI summaries in the benchtop comparison.

\begin{figure}[!htbp]
  \centering
  \includegraphics[width=0.8\linewidth]{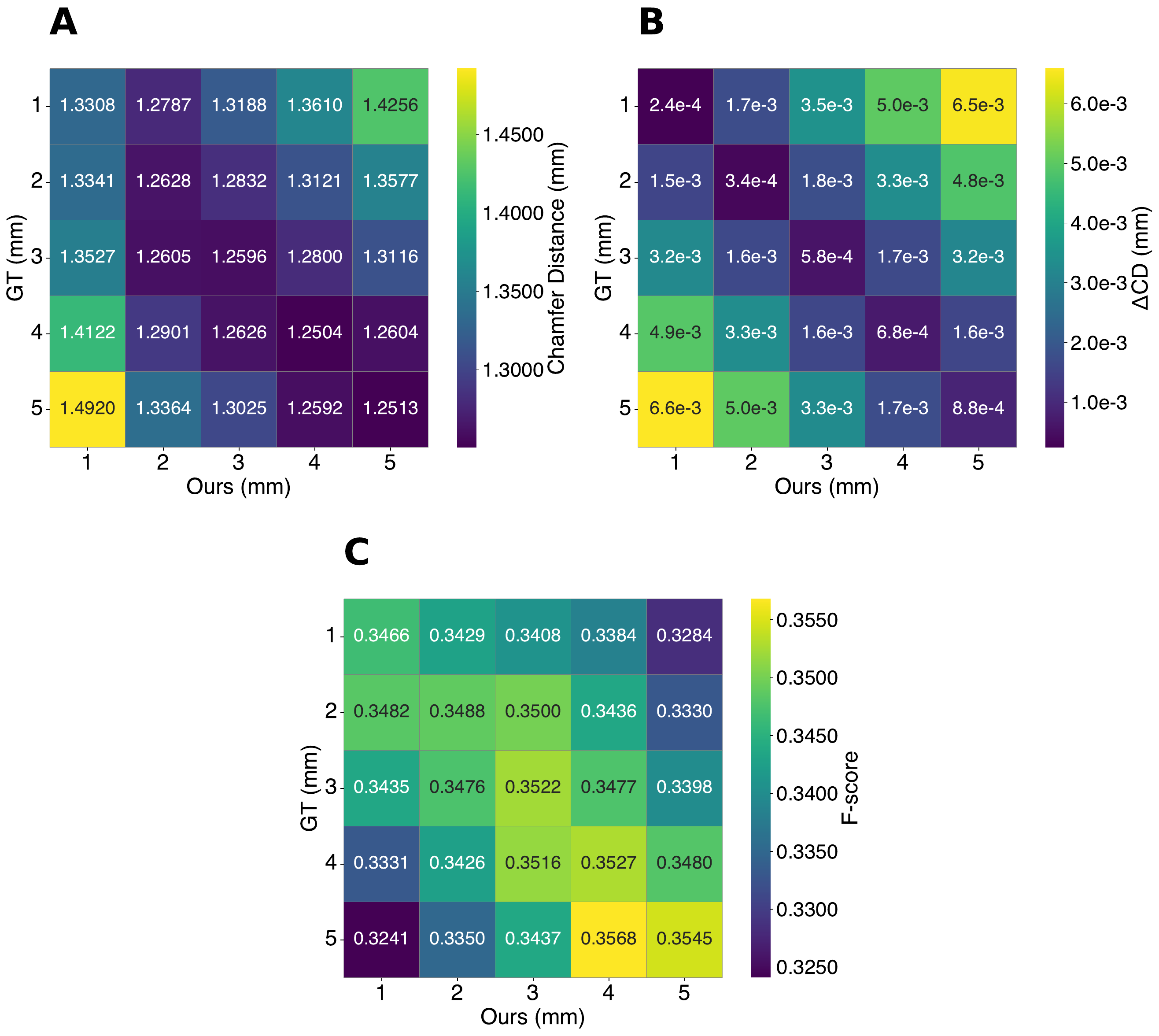}
  \caption{GT-versus-reconstruction agreement across synthetic pull magnitudes (condition-level comparison; not a single-frame plot). Pairwise heatmaps report (A) Chamfer distance (CD, mm), (B) temporal Chamfer delta ($\Delta\mathrm{CD}$, mm), and (C) F-score; metric definitions follow Sec.~\ref{sec:synthetic_validation_metrics} (Eqs.~\ref{eq:chamfer}, \ref{eq:delta_cd_main}, and \ref{eq:fscore_main}). Best agreement occurs along the diagonal (matched pull conditions).}
  \label{fig:synth_heatmaps}
\end{figure}

\begin{table}[!htbp]
  \centering
  \caption{Synthetic validation metrics by deformation condition. Definitions for CD, $\mathrm{CD}_{\text{norm}}$, $\Delta\mathrm{CD}_{\text{rel}}$, precision, recall, and F-score follow Sec.~\ref{sec:synthetic_validation_metrics} (Eqs.~\ref{eq:chamfer}, \ref{eq:delta_cd_main}, \ref{eq:pr_main}, and \ref{eq:fscore_main}). Displacement error (mm, \%) and stress bias (MPa) are reported as Ours $-$ GT using the max-median ROI definitions described in Sec.~\ref{sec:ROI_metrics}.}
  \label{tab:synthetic_validation}
  \vspace{0.4cm}
  \resizebox{\textwidth}{!}{%
    \begin{tabular}{lcccccccc}
      \toprule
      \textbf{Condition} & \multicolumn{2}{c}{\textbf{CD}} & \textbf{$\Delta\mathrm{CD}_{\text{rel}}$} & \textbf{Precision} & \multicolumn{2}{c}{\textbf{Displacement Error}} & \textbf{Stress Bias} \\
      \cmidrule(lr){2-3} \cmidrule(lr){6-7}
      & \textbf{(mm)} & \textbf{$\mathrm{CD}_{\text{norm}}$} & & & \textbf{(mm)} & \textbf{(\%)} & \textbf{(MPa)} \\
      \midrule
      Bulk & 1.881 & 3.152 & 0.085 & 0.919 & 0.091 & 0.64 & 0.018 \\
      1 mm & 1.815 & 3.041 & 0.139 & 0.970 & 0.047 & 7.17 & 0.004 \\
      2 mm & 1.752 & 2.937 & 0.100 & 0.967 & 0.094 & 7.82 & 0.004 \\
      3 mm & 1.738 & 2.912 & 0.114 & 0.964 & 0.087 & 4.98 & 0.007 \\
      4 mm & 1.745 & 2.924 & 0.100 & 0.972 & 0.164 & 7.38 & 0.007 \\
      5 mm & 1.714 & 2.872 & 0.105 & 0.968 & 0.190 & 5.28 & 0.006 \\
      \bottomrule
    \end{tabular}%
  }
\end{table}

\subsection*{Benchtop comparative application}
\begin{figure}[!htbp]
  \centering
  \includegraphics[width=0.6\linewidth]{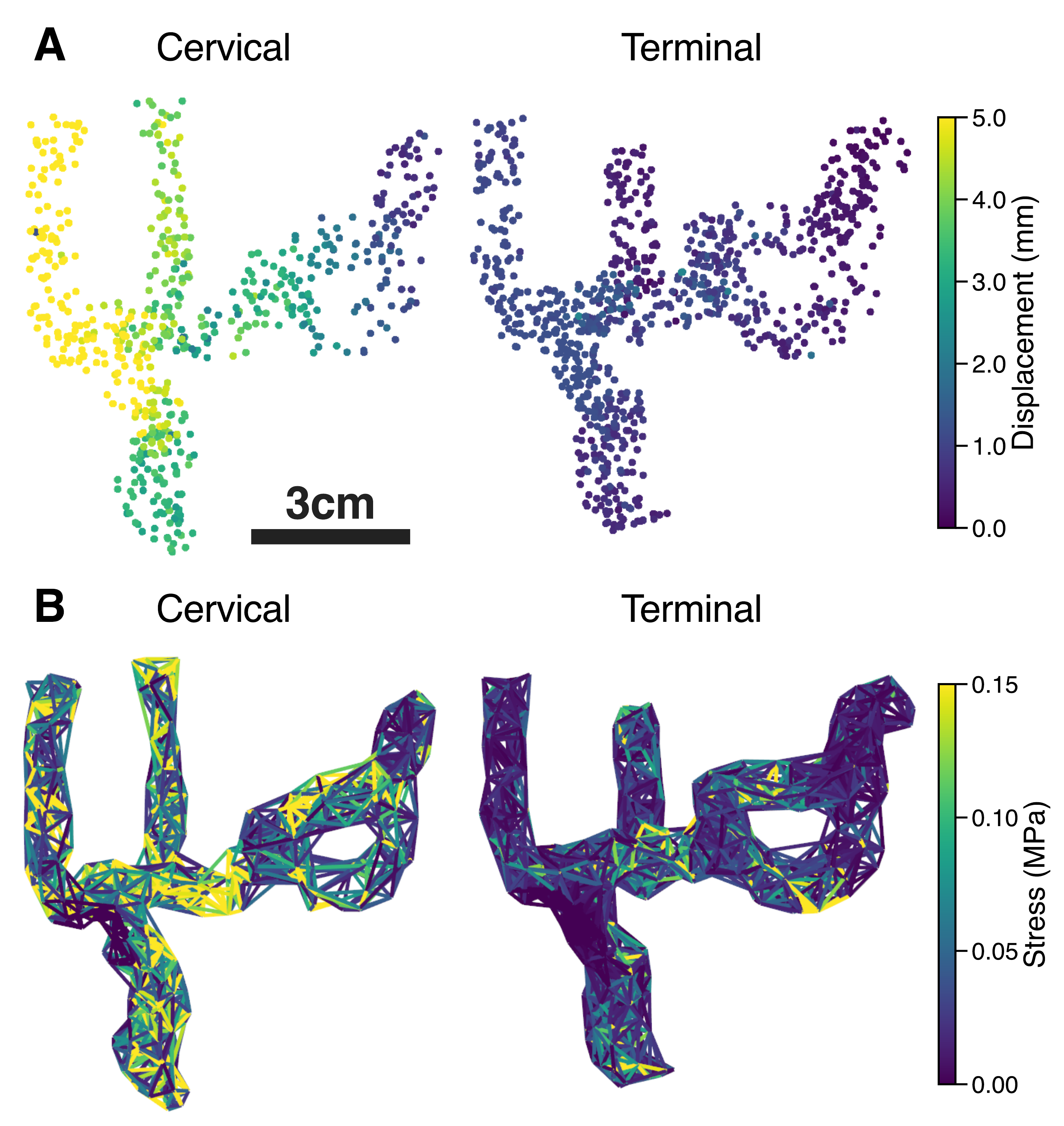}
  \caption{Benchtop application demonstration at peak deformation for a single representative trial per AC placement condition (cervical and terminal). (A) Reconstructed displacement magnitude (mm) relative to the initial frame. (B) Corresponding edge-based stress-proxy magnitude $|\sigma_e|$ (MPa) at the same peak-deformation timepoint.}
  \label{fig:results_conditions}
\end{figure}

Using the same validated metrics, the benchtop comparison (one trial per condition) showed broader and larger displacement/stress-proxy fields for AC cervical placement than ICA terminus placement (Fig.~\ref{fig:results_conditions}).

\begin{figure}[!htbp]
  \centering
  \includegraphics[width=\linewidth]{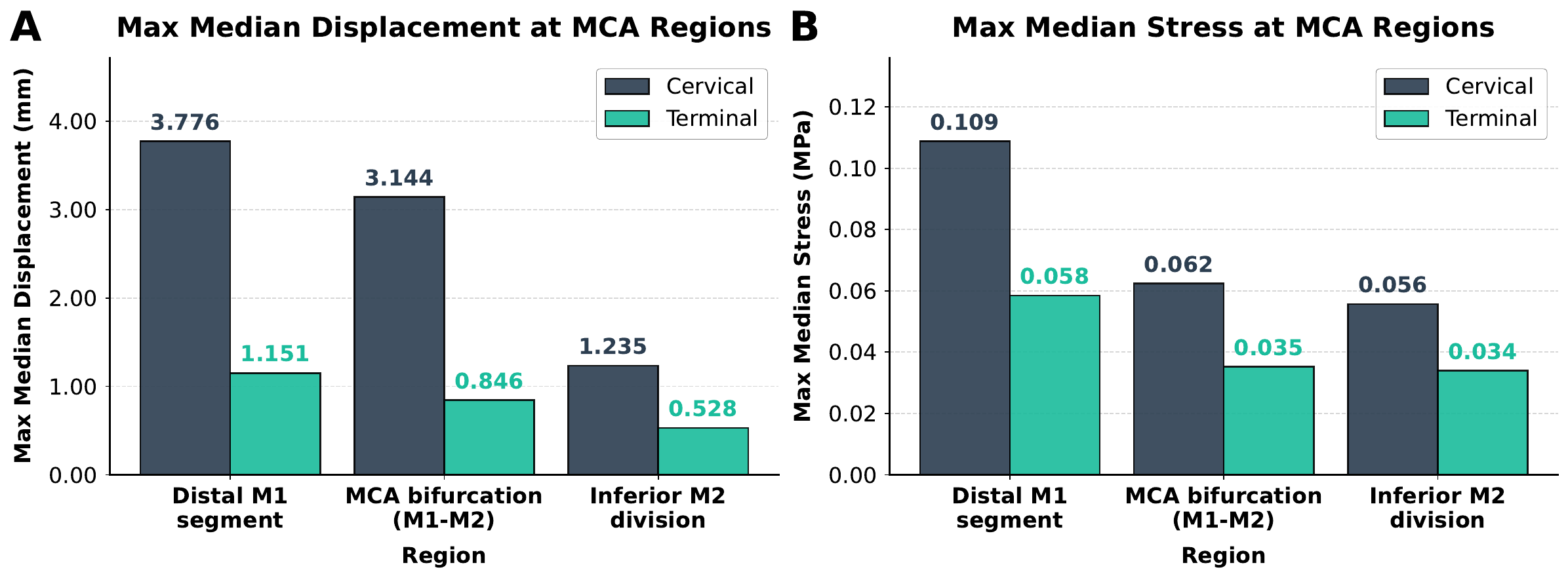}
  \caption{Regional max-median displacement and stress-proxy during thrombectomy for a single representative trial per AC placement condition. Bars report the maximum over time of the regional median displacement magnitude (A, mm) and edge-based stress-proxy magnitude (B, MPa) across distal M1 segment, MCA bifurcation (M1/M2), and inferior M2 division.}
  \label{fig:results_regional_bars}
\end{figure}

Max-median ROI values (Fig.~\ref{fig:results_regional_bars}) were higher in cervical vs terminal placement for displacement: distal M1 segment 3.776 vs 1.151~mm, MCA bifurcation (M1/M2) 3.144 vs 0.846~mm, and inferior M2 division 1.235 vs 0.528~mm. Max-median stress proxy was also higher in cervical vs terminal placement: distal M1 segment 0.109 vs 0.058~MPa, bifurcation 0.062 vs 0.035~MPa, and inferior M2 division 0.056 vs 0.034~MPa. Corresponding benchtop videos are provided in \VideoRef{S2}, \VideoRef{S3}, \VideoRef{S4}, and \VideoRef{S5}. Given single-trial sampling, these values are interpreted as comparative directional observations.

\subsection*{Scope of reported quantities}
Consistent with the methods scope, stress values are reported as relative surface-based stress-proxy magnitudes (\(|\sigma_e|\)) from reconstructed surface deformation. They are unsigned comparative metrics and do not represent absolute wall stress or tensile/compressive sign. The rigid-motion control supports this interpretation by showing near-zero stress proxy under bulk translation for most edges.

\FloatBarrier

\section{Discussion}
This methods/validation study defines and tests a multi-view reconstruction protocol for dynamic benchtop thrombectomy experiments, moving from workflow definition to synthetic validation and then benchtop comparison. Three findings support the utility of the approach. First, the rigid-motion control produced near-zero stress proxy for most edges, indicating expected suppression of spurious internal loading under bulk translation. Second, synthetic pulling experiments showed moderate absolute geometric error but strong temporal and task-relevant agreement across conditions, supporting use of the framework for comparative kinematic analysis. Third, the benchtop comparison showed a consistent directional difference between cervical ICA and ICA terminus AC placement in both max-median displacement and max-median stress-proxy outputs.

These trends are broadly consistent with prior work showing that thrombectomy strategy and device configuration influence vessel loading and procedural risk. In vitro evidence has shown that combined stent-retriever and aspiration-catheter techniques can reduce vessel stretching relative to stent retriever alone in selected MCA configurations \citep{nariaiModificationAspirationCatheter2023}. Clinical studies likewise suggest tradeoffs between distal access strategy, reperfusion performance, and vascular injury: more distal access catheter tip position can improve reperfusion but has also been associated with higher vascular injury rates \citep{baekEffectDistalAccess2021}, and recent cohort analyses have linked device/strategy factors to recanalization performance \citep{kawamotoStentretrieverCharacteristicsStrategies2024}. Relative to prior benchtop studies that emphasize removal forces, friction, or sparse landmark-based motion surrogates \citep{poulosInvestigationStentRetriever2024,nagargojeRoleFrictionForces2025,nariaiModificationAspirationCatheter2023}, the present framework contributes time-resolved, spatially resolved vessel-surface kinematics and regional comparative loading maps at anatomically defined ROIs.

The benchtop comparison is intended as a comparative demonstration rather than an inferential biomechanical conclusion because each condition includes one representative trial. Likewise, the synthetic validation is designed to test whether the reconstruction-and-analysis pipeline recovers known imposed surface motion, preserves temporal behavior, and suppresses spurious stress-proxy response under rigid translation; it does not validate absolute biomechanical realism, device--clot--vessel interaction mechanics, or the constitutive fidelity of the stress mapping. It also does not test the video-segmentation stage under realistic benchtop imaging conditions, because the Blender renders use transparent backgrounds rather than the cluttered backgrounds present in the physical videos that must be filtered with the SAM~2-based masking workflow. In addition, the synthetic vessel appearance may not fully match the optical surface characteristics of the real phantom, including vessel-surface contrast and the apparent size or visibility of the surface beads, so reconstruction behavior in the synthetic setting may differ from performance on benchtop footage. The stress proxy is derived from reconstructed surface geometry and should therefore be interpreted as a relative surface-based measure rather than an absolute wall-stress estimate. It does not resolve through-wall mechanics, lumen-facing stresses, or stress sign, and the edge-based Neo-Hookean mapping is a simplified constitutive surrogate rather than a full continuum model. More direct estimation of device-, clot-, and vessel-level internal mechanics would require complementary continuum or coupled in-silico modeling approaches \citep{luraghiApplicabilityAssessmentStentretriever2021,mousavijsRealisticComputerModelling2021}. Experiments were performed in silicone phantoms, which are valuable for controlled testing but do not fully reproduce in vivo tissue properties, artery--clot--device interaction, or physiological boundary conditions \citep{liuPreclinicalTestingPlatforms2021,johnsonReviewAdvancementsInvitro2022,luraghiApplicabilityAssessmentStentretriever2021}.

Future work should therefore include repeated trials across a broader range of anatomies, clot analogs, and device configurations, together with cross-validation against complementary experimental or computational models. Methodologically, an important next step would be to move from the current fixed post hoc edge-graph representation toward more explicit surface-aware dynamic Gaussian formulations. Static work such as SuGaR motivates surface-aligned mesh extraction from Gaussian representations, while more recent dynamic methods such as DynaSurfGS, DG-Mesh, and GSTAR suggest possible paths toward temporally consistent surface reconstruction and tracking directly within Gaussian-based pipelines \citep{guedonSuGaRSurfaceAlignedGaussian2023a,caiDynaSurfGSDynamicSurface2024,liuDynamicGaussiansMesh2025,zhengGauSTARGaussianSurface2025}. If future extensions move beyond comparative kinematics toward coupled reconstruction-and-simulation workflows, mesh-coupled Gaussian methods such as MaGS may also be relevant \citep{maMaGSReconstructingSimulating2024}. On the segmentation side, beyond the current SAM 2 workflow, future work could evaluate motion-aware and memory-based video segmentation approaches such as SAMURAI, Cutie, and DEVA to reduce manual correction and improve temporal consistency under self-occlusion, distractors, or longer sequences; concept-promptable models such as SAM 3 may also become useful if later datasets require more flexible open-vocabulary prompting \citep{yangSAMURAIAdaptingSegment2024,chengPuttingObjectBack2024,chengTrackingAnythingDecoupled2023,carionSAM3Segment2025}.

\section{Conclusion}
This work introduces a standardized and accessible protocol for comparative benchtop thrombectomy analysis by combining low-cost multi-view acquisition, dynamic 4D reconstruction, fixed-connectivity ROI tracking, and relative surface-based stress-proxy mapping. The main novelty is the integration of these components with synthetic ground-truth validation and a clinically relevant two-condition benchtop demonstration in a single workflow, which together provide a coherent acquisition-to-analysis framework.

The framework provides practical, time-resolved regional metrics that can support method development and hypothesis generation for thrombectomy-device and access-strategy testing, while maintaining clear separation from absolute biomechanical wall-stress inference. Future work will focus on repeated trials per condition, expanded aspiration-catheter and pullback-condition sampling, broader anatomical variability, and improved reconstruction robustness through better camera coverage, reduced self-occlusion, and stronger temporal regularization.

\ifArxivFormat
\section*{Acknowledgments}
This work was supported by the Elegant Mind Collaboration, the UCLA Department of Physics and Astronomy, and the Kaneko Lab at UCLA Health.

\medskip
\noindent\textbf{Author contributions (CRediT).}

\textbf{Ethan Nguyen}: Conceptualization, Methodology, Software, Resources, Data Curation, Supervision, Formal Analysis, Investigation, Validation, Visualization, Project Administration, Writing -- original draft, Writing -- review \& editing.

\textbf{Javier Carmona}: Conceptualization, Methodology, Software, Investigation, Supervision, Writing -- review \& editing.

\textbf{Arisa Matsuzaki}: Investigation, Validation, Resources, Writing -- review \& editing.

\textbf{Naoki Kaneko}: Conceptualization, Resources, Funding Acquisition, Supervision, Writing -- review \& editing.

\textbf{Katsushi Arisaka}: Conceptualization, Funding Acquisition, Supervision, Writing -- review \& editing.

\noindent All authors reviewed and approved the final manuscript.

\medskip
The authors thank members of the Elegant Mind Collaboration for their support. Soumya Bukkapatnam assisted with manual placement of fluorescent beads on the vessel models. Charlie Long designed the initial iterations of the camera mounts. Thomas Leung and Martin Leung constructed the dodecahedron camera rig. Mingda He assisted with code development.

The authors also thank members of the Kaneko Lab. Dr. Eisuke Tsukagoshi and Lea Guo provided the silicone vessel models. Lea Guo and Mahsa Ghovvati also assisted with equipment procurement.

The authors further thank Dr. Shusaku Goto (visiting from Nagoya Hospital), Dr. Natsuhi Sasaki (research scholar at UCLA), and Dr. Eisuke Tsukagoshi for conducting the physical thrombectomy experiments.

\section*{Preprint Notes}

\noindent \textbf{Version:} Version 1 April 2026.

\noindent \textbf{Preprint status:} This manuscript is a preprint and has not yet undergone peer review. It is intended for community feedback.

\noindent \textbf{Funding:} This work was supported by UCLA Physics \& Astronomy / UCLA Health.

\noindent \textbf{Ethics statement:} This study did not involve human subjects or animal experiments.

\noindent \textbf{Data availability:} Code and example data are publicly available at \url{https://ethanuser.github.io/vessel4D}. Additional benchtop data are available from the corresponding authors upon reasonable request.

\noindent \textbf{Conflicts of interest:} The authors declare no conflicts of interest.
\else
\section*{Acknowledgments}
Removed for anonymity during peer review.

\section*{Preprint Notes}

\noindent \textbf{Version:} Version 1 April 2026.

\noindent \textbf{Funding:} Removed for anonymity during peer review.

\noindent \textbf{Ethics statement:} This study did not involve human subjects or animal experiments.

\noindent \textbf{Data availability:} Removed for anonymity during peer review.

\noindent \textbf{Conflicts of interest:} The authors declare no conflicts of interest.
\fi

\ifArxivFormat
\bibliographystyle{elsarticle-num}
\else
\bibliographystyle{plainnat}
\fi
\bibliography{refs}

\ifArxivFormat
\setcounter{table}{0}
\renewcommand{\thetable}{S\arabic{table}}
\setcounter{figure}{0}
\renewcommand{\thefigure}{S\arabic{figure}}
\setcounter{section}{0}
\renewcommand{\thesection}{S\arabic{section}}
\renewcommand{\thesubsection}{S\arabic{section}.\arabic{subsection}}
\renewcommand{\theHsection}{supp.S\arabic{section}}
\renewcommand{\theHsubsection}{supp.S\arabic{section}.\arabic{subsection}}
\renewcommand{\theHfigure}{supp.S\arabic{figure}}
\renewcommand{\theHtable}{supp.S\arabic{table}}
\setcounter{secnumdepth}{2}

\section*{Supplementary Information}
This supplement provides overflow material that supports, but is not required to understand, reproduce, or review the main methods/validation framework.

\section{Supplementary Methods}
This section retains only implementation details and supporting tables/media references that are not needed in the main manuscript.

\subsection{Acquisition implementation details}
For reproducibility, the incoming OBS composite was recorded at 4080\(\times\)4080 resolution using the NVENC H.264 encoder in \texttt{.mkv} format with YUY2 color, ``High Quality, Medium File Size'' recording quality, white balance set to approximately 2498, maximum exposure set to \(-11\), and an OBS focus setting of approximately 500--525. A single UV lamp was positioned beneath the apparatus, and the camera/lighting settings were held fixed across experiments. No formal temporal synchronization measurement beyond visual inspection was performed; in practice, the light source was toggled before recording to confirm that all cameras were awake and actively responding.

\subsection{Segmentation and graph-construction implementation details}
The 4080\(\times\)4080 OBS composite was first split into nine 1360\(\times\)1360 view videos, and SAM~2 was run independently on each view before the segmented outputs were merged back into the multi-view reconstruction workflow. Typical segmentation used approximately eight additive/subtractive prompt points per vessel. Common failure modes were view-specific occlusion of the vessel by parts of the 3D-printed support structure and reduced fluorescent-bead visibility near the vessel ends; manual re-prompting was uncommon (less than 20\% of sequences), and no mask cleanup beyond background suppression was applied.

Before clustering and graph construction, primitives were removed if their estimated radius was below \SI{0.07}{mm}, if opacity was below 0.05, or if their RGB standard deviation was below 0.05 (used to reject near-gray/background-like points). Reconstruction otherwise followed the default 4DGS settings summarized in the main manuscript. Typical runs retained on the order of \(\sim\)70{,}000 filtered primitives, which were reduced to roughly 2{,}000--3{,}000 clusters after clustering and to roughly \(\sim\)2{,}000 curated clusters after manual cleanup; the corresponding graph typically contained on the order of 50{,}000--60{,}000 edges. Formal clustering, graph, spatial-coherence, and metric definitions remain in the main manuscript.

\section{Synthetic Validation}
This section adds condition-level metrics and brief implementation details that complement the main-text synthetic summary. Table~\ref{tab:synthetic_validation_supp} reports the expanded condition-level metrics. In Blender, the bulk control used a rigid translation, whereas the pulling conditions used a lattice deformation in which the main control point was displaced by \(n\) mm for the nominal ``\(n\) mm'' pull.

\noindent Metric definitions follow the main manuscript: CD (Eq.~\ref{eq:chamfer}); directed distances \(d_{P\to G}\) and \(d_{G\to P}\) (Eqs.~\ref{eq:directed_dist_p2g} and \ref{eq:directed_dist_g2p}); temporal disagreement \(\Delta\mathrm{CD}\) (Eq.~\ref{eq:delta_cd_main}); precision/recall (Eq.~\ref{eq:pr_main}); and F-score (Eq.~\ref{eq:fscore_main}).

\begin{table}[htbp]
  \centering
  \caption{Synthetic validation metrics across deformation conditions, including directed and symmetric Chamfer terms, temporal disagreement, and overlap statistics.}
  \label{tab:synthetic_validation_supp}
  \vspace{0.4cm}
  \resizebox{\textwidth}{!}{%
    \begin{tabular}{lcccccccccc}
      \toprule
      \multirow{2}{*}{\textbf{Condition}} & \multicolumn{4}{c}{\textbf{CD}} & \multicolumn{2}{c}{\textbf{$\Delta\mathrm{CD}$}} & \multirow{2}{*}{\textbf{$\tau$ (mm)}} & \multirow{2}{*}{\textbf{P}} & \multirow{2}{*}{\textbf{R}} & \multirow{2}{*}{\textbf{F}} \\
      \cmidrule(lr){2-5} \cmidrule(lr){6-7}
      & \textbf{$d_{P\to G}$ (mm)} & \textbf{$d_{G\to P}$ (mm)} & \textbf{Sym (mm)} & \textbf{$\mathrm{CD}_{\text{norm}}$} & \textbf{mm} & \textbf{rel} & & & & \\
      \midrule
      \multirow{3}{*}{Bulk} & \multirow{3}{*}{0.535} & \multirow{3}{*}{1.345} & \multirow{3}{*}{1.881} & \multirow{3}{*}{3.152} & \multirow{3}{*}{$1.52 \times 10^{-2}$} & \multirow{3}{*}{0.085} & 1 & 0.919 & 0.313 & 0.467 \\
      & & & & & & & 2 & 0.983 & 0.874 & 0.925 \\
      & & & & & & & 3 & 0.998 & 0.981 & 0.989 \\
      \midrule
      \multirow{3}{*}{1 mm} & \multirow{3}{*}{0.465} & \multirow{3}{*}{1.349} & \multirow{3}{*}{1.815} & \multirow{3}{*}{3.041} & \multirow{3}{*}{$4.75 \times 10^{-4}$} & \multirow{3}{*}{0.139} & 1 & 0.970 & 0.309 & 0.468 \\
      & & & & & & & 2 & 0.998 & 0.863 & 0.926 \\
      & & & & & & & 3 & 0.999 & 0.981 & 0.990 \\
      \midrule
      \multirow{3}{*}{2 mm} & \multirow{3}{*}{0.474} & \multirow{3}{*}{1.279} & \multirow{3}{*}{1.752} & \multirow{3}{*}{2.937} & \multirow{3}{*}{$6.82 \times 10^{-4}$} & \multirow{3}{*}{0.100} & 1 & 0.967 & 0.344 & 0.507 \\
      & & & & & & & 2 & 0.995 & 0.890 & 0.940 \\
      & & & & & & & 3 & 0.999 & 0.991 & 0.995 \\
      \midrule
      \multirow{3}{*}{3 mm} & \multirow{3}{*}{0.462} & \multirow{3}{*}{1.276} & \multirow{3}{*}{1.738} & \multirow{3}{*}{2.912} & \multirow{3}{*}{$1.16 \times 10^{-3}$} & \multirow{3}{*}{0.114} & 1 & 0.964 & 0.348 & 0.512 \\
      & & & & & & & 2 & 0.994 & 0.891 & 0.940 \\
      & & & & & & & 3 & 0.998 & 0.989 & 0.994 \\
      \midrule
      \multirow{3}{*}{4 mm} & \multirow{3}{*}{0.465} & \multirow{3}{*}{1.279} & \multirow{3}{*}{1.745} & \multirow{3}{*}{2.924} & \multirow{3}{*}{$1.37 \times 10^{-3}$} & \multirow{3}{*}{0.100} & 1 & 0.972 & 0.346 & 0.510 \\
      & & & & & & & 2 & 0.997 & 0.889 & 0.940 \\
      & & & & & & & 3 & 0.998 & 0.990 & 0.994 \\
      \midrule
      \multirow{3}{*}{5 mm} & \multirow{3}{*}{0.465} & \multirow{3}{*}{1.249} & \multirow{3}{*}{1.714} & \multirow{3}{*}{2.872} & \multirow{3}{*}{$1.79 \times 10^{-3}$} & \multirow{3}{*}{0.105} & 1 & 0.968 & 0.356 & 0.521 \\
      & & & & & & & 2 & 0.995 & 0.905 & 0.948 \\
      & & & & & & & 3 & 0.999 & 0.993 & 0.996 \\
      \bottomrule
    \end{tabular}%
  }
\end{table}

\noindent At the default \(\tau=\SI{1}{mm}\) threshold, recall is lower than precision because the reconstruction is summarized as a clustered point cloud that is intentionally sparser than the dense GT sampling. In other words, most reconstructed points still lie near the GT surface, but many GT points do not have a reconstructed point within the same tolerance because clustering reduces point density to improve temporal stability and suppress local Gaussian drift/noise. This asymmetry is why precision remains high while recall and F-score are more conservative in Table~\ref{tab:synthetic_validation_supp}.

\section{Additional Quantitative Results}
Table~\ref{tab:regional_summary} reports regional median displacement and stress values across sweep settings for ground truth and reconstructed estimates, using the ROI displacement/stress definitions in Sec.~\ref{sec:ROI_metrics} of the main manuscript (including Eqs.~\ref{eq:roi_displacement}--\ref{eq:neohookean_uniaxial}). Percent error is intentionally omitted here, because small ground-truth magnitudes can inflate relative percentage error and obscure practical agreement. Agreement is assessed primarily through Bland--Altman analysis and correlation trends in the main manuscript.

\begin{table}[htbp]
  \centering
  \caption{Regional summary statistics (median) for displacement and stress across sweep settings. Agreement is evaluated primarily via Bland--Altman and correlation in the main manuscript.}
  \label{tab:regional_summary}
  \vspace{0.3cm}
  \resizebox{\textwidth}{!}{%
    \begin{tabular}{llcccccccccc}
      \toprule
      \multirow{2}{*}{\textbf{Region}} & \multirow{2}{*}{\textbf{Method}} & \multicolumn{5}{c}{\textbf{Displacement (mm)}} & \multicolumn{5}{c}{\textbf{Stress (MPa)}} \\
      \cmidrule(lr){3-7} \cmidrule(lr){8-12}
      & & \textbf{1 mm} & \textbf{2 mm} & \textbf{3 mm} & \textbf{4 mm} & \textbf{5 mm} & \textbf{1 mm} & \textbf{2 mm} & \textbf{3 mm} & \textbf{4 mm} & \textbf{5 mm} \\
      \midrule
      \multirow{2}{*}{\textbf{R1}} & GT  & 1.023 & 2.046 & 3.068 & 4.091 & 5.114 & 0.005 & 0.009 & 0.014 & 0.018 & 0.023 \\
      & Ours  & 0.999 & 2.040 & 3.074 & 4.116 & 5.079 & 0.006 & 0.010 & 0.016 & 0.017 & 0.022 \\ \midrule
      \multirow{2}{*}{\textbf{R2}} & GT  & 0.956 & 1.912 & 2.868 & 3.823 & 4.779 & 0.007 & 0.015 & 0.022 & 0.029 & 0.036 \\
      & Ours  & 0.922 & 1.925 & 2.791 & 3.731 & 4.750 & 0.009 & 0.015 & 0.025 & 0.037 & 0.034 \\ \midrule
      \multirow{2}{*}{\textbf{R3}} & GT  & 0.302 & 0.604 & 0.906 & 1.208 & 1.510 & 0.010 & 0.021 & 0.031 & 0.042 & 0.052 \\
      & Ours  & 0.275 & 0.507 & 0.789 & 1.004 & 1.458 & 0.015 & 0.023 & 0.038 & 0.045 & 0.058 \\ \midrule
      \multirow{2}{*}{\textbf{R4}} & GT  & 0.596 & 1.193 & 1.789 & 2.385 & 2.981 & 0.010 & 0.022 & 0.034 & 0.046 & 0.059 \\
      & Ours  & 0.639 & 1.114 & 1.847 & 2.416 & 2.999 & 0.016 & 0.026 & 0.040 & 0.052 & 0.064 \\ \midrule
      \multirow{2}{*}{\textbf{R5}} & GT  & 0.821 & 1.642 & 2.463 & 3.284 & 4.105 & 0.008 & 0.016 & 0.024 & 0.033 & 0.041 \\
      & Ours  & 0.724 & 1.400 & 2.250 & 3.021 & 3.507 & 0.012 & 0.021 & 0.033 & 0.042 & 0.052 \\
      \bottomrule
    \end{tabular}%
  }
\end{table}

\subsection{Spatial coherence filtering ablation}
To assess the effect of spatial coherence filtering (SCF), we repeated the synthetic ROI agreement analysis without SCF using the same 25 paired points (5 ROIs $\times$ 5 pull magnitudes). Figure~\ref{fig:supp_agreement_no_scf} shows the corresponding correlation and Bland--Altman plots, and Fig.~\ref{fig:supp_scf_compare_qual} provides a qualitative stress-map comparison for the last frame (maximum deformation) of the 5~mm pulling condition. Without SCF, stray high-stress points are more visibly apparent in localized regions, whereas SCF suppresses these artifacts while preserving the broader stress pattern.

\begin{figure}[ht]
  \centering
  \includegraphics[width=0.8\linewidth]{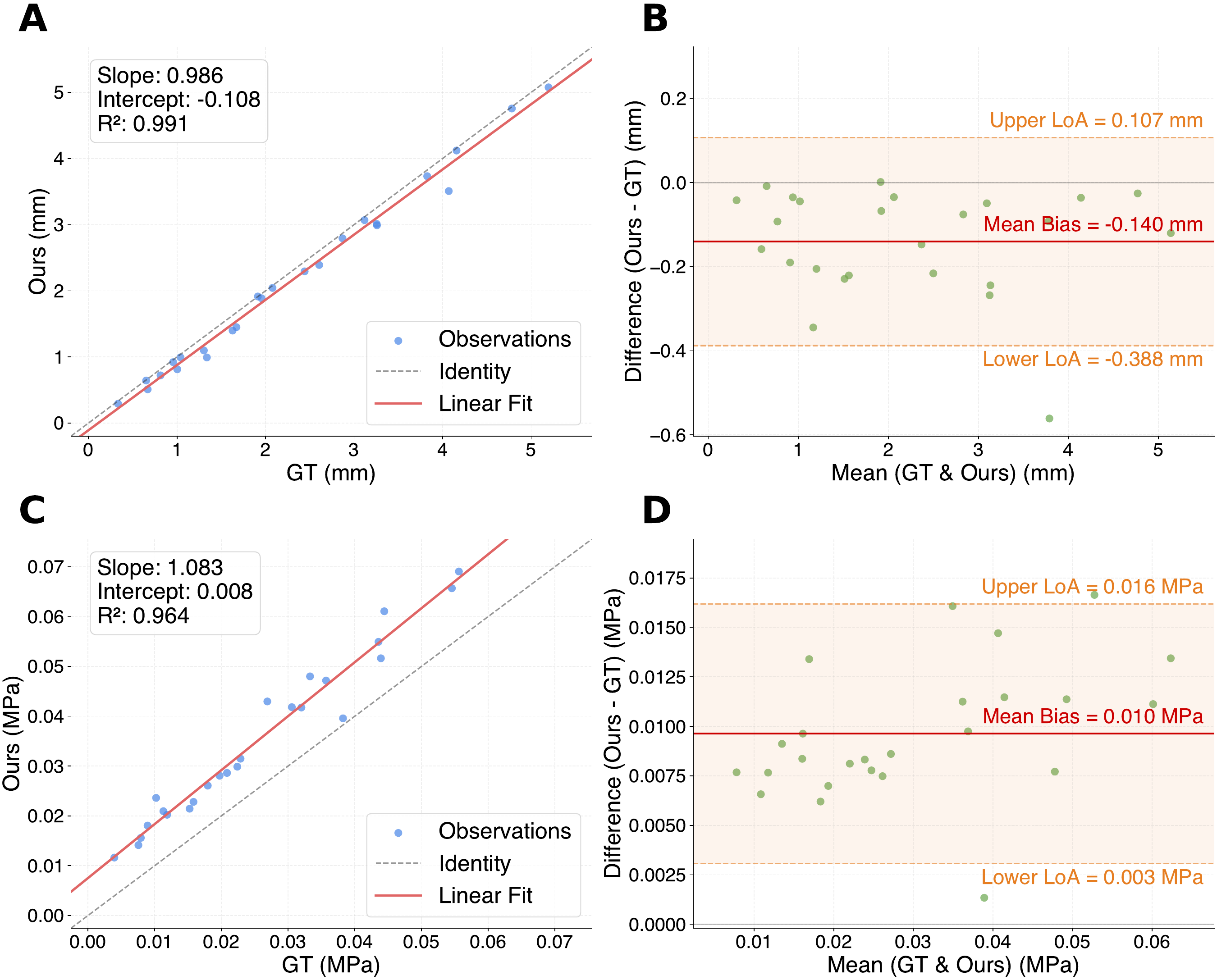}
  \caption{Agreement analyses without spatial coherence filtering (SCF) for max-median ROI metrics between reconstruction and ground truth: correlation and Bland--Altman for displacement magnitude (A--B, mm) and relative surface-based stress-proxy magnitude (C--D, MPa).}
  \label{fig:supp_agreement_no_scf}
\end{figure}

\begin{figure}[!htbp]
  \centering
  \includegraphics[width=\linewidth]{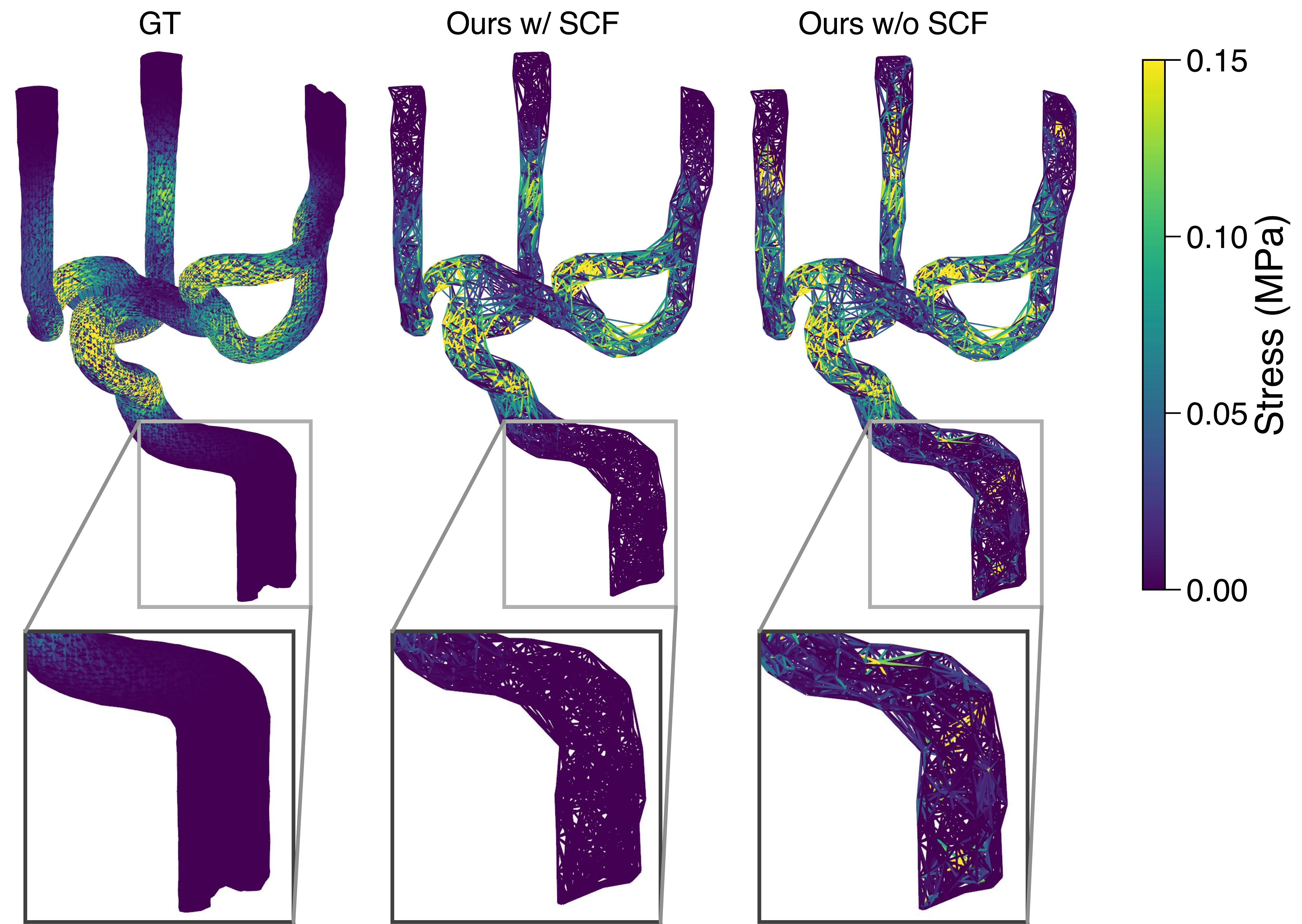}
  \caption{Qualitative stress-proxy comparison with and without spatial coherence filtering (SCF) at the last frame (maximum deformation) of the 5~mm synthetic pulling condition. GT, reconstruction with SCF, and reconstruction without SCF are shown with a shared color scale. The zoomed view shows more visible stray high-stress points without SCF, whereas SCF suppresses these localized artifacts while preserving the overall stress pattern.}
  \label{fig:supp_scf_compare_qual}
\end{figure}

\begin{table}[htbp]
  \centering
  \caption{Agreement summary with and without spatial coherence filtering (SCF) for max-median ROI synthetic validation metrics.}
  \label{tab:scf_ablation_compare}
  \vspace{0.3cm}
  \resizebox{\textwidth}{!}{%
    \begin{tabular}{llcccccc}
      \toprule
      \textbf{Metric} & \textbf{Condition} & \textbf{Slope} & \textbf{Intercept} & \textbf{$R^2$} & \textbf{BA Bias} & \textbf{Lower LoA} & \textbf{Upper LoA} \\
      \midrule
      \multirow{2}{*}{Displacement (mm)} & w/ SCF & 0.987 & -0.112 & 0.992 & -0.142 & -0.385 & 0.100 \\
      & w/o SCF & 0.986 & -0.108 & 0.991 & -0.140 & -0.388 & 0.107 \\
      \midrule
      \multirow{2}{*}{Stress proxy (MPa)} & w/ SCF & 1.068 & 0.002 & 0.969 & 0.004 & -0.002 & 0.010 \\
      & w/o SCF & 1.083 & 0.008 & 0.964 & 0.010 & 0.003 & 0.016 \\
      \bottomrule
    \end{tabular}%
  }
\end{table}

\noindent Relative to the SCF-enabled analysis, displacement agreement metrics remain similar, while stress-proxy agreement shows larger positive bias and wider limits of agreement without SCF. Specifically, stress-proxy Bland--Altman bias increases from \SI{0.004}{MPa} to \SI{0.010}{MPa}, and the upper limit of agreement widens from \SI{0.010}{MPa} to \SI{0.016}{MPa}; qualitatively, the no-SCF maps also show more localized stray high-stress points (Fig.~\ref{fig:supp_scf_compare_qual}).

\section{Supplementary Figures}
\begin{figure}[!htbp]
  \centering
  \includegraphics[width=\linewidth]{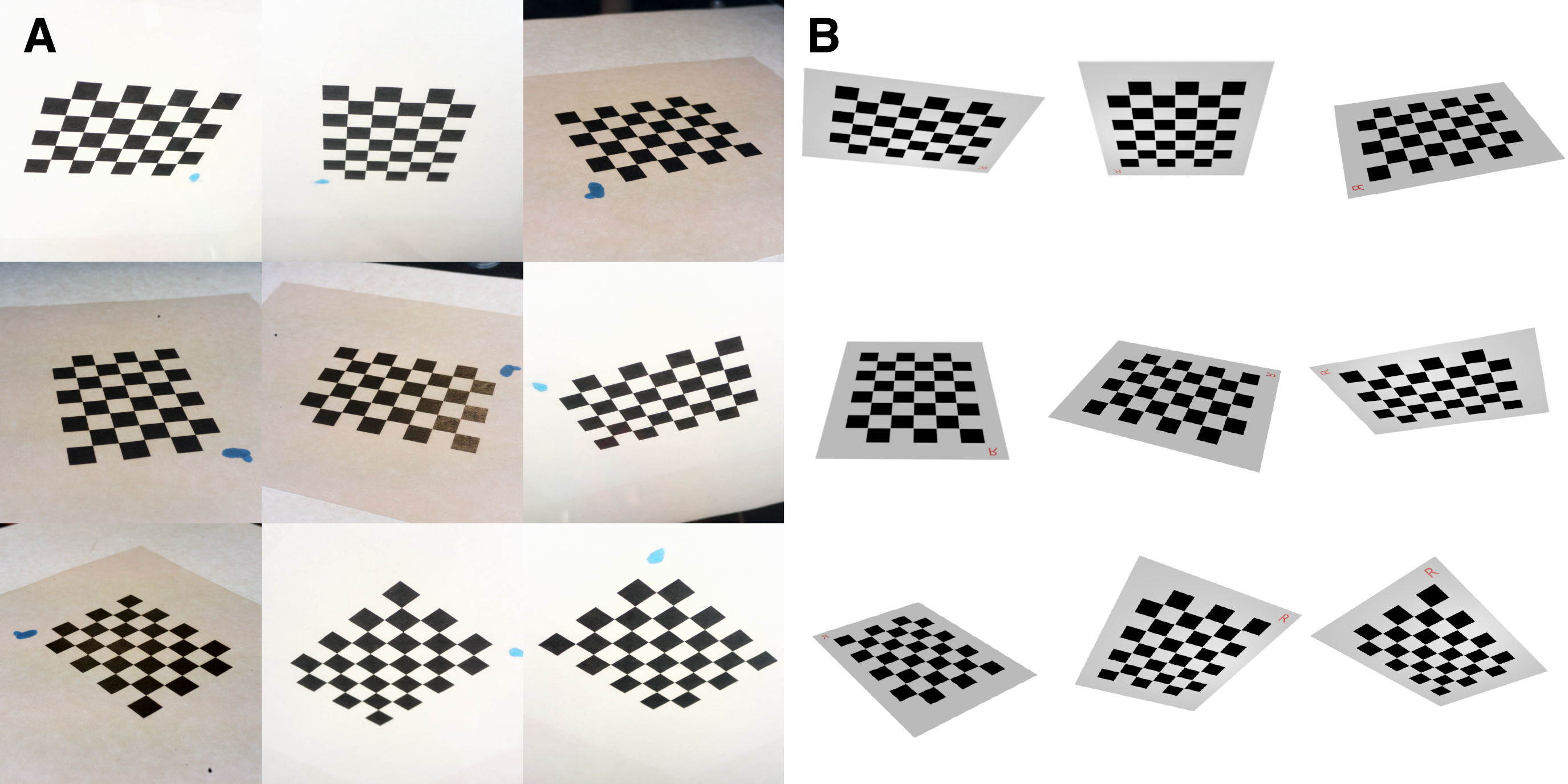}
  \caption{Nine-view chessboard calibration comparison between physical and synthetic setups. (A) Real-camera views from the nine-camera rig using the \(6\times8\) chessboard used to estimate camera extrinsic parameters. (B) Corresponding views from the nine synthetic Blender cameras using the same pattern.}
  \label{fig:supp_calibration_pattern}
\end{figure}

\FloatBarrier
\clearpage

\noindent
\begin{minipage}{\linewidth}
  \section{Supplementary Videos and Data}
  Table~\ref{tab:supp_videos} indexes all supplementary videos using labels S1--S11 and provides a short description for each item.

  \centering
  \captionof{table}{Supplementary video index and short descriptions.}
  \label{tab:supp_videos}
  \vspace{0.3cm}
  \resizebox{\textwidth}{!}{%
    \begin{tabular}{cllp{8.5cm}c}
      \toprule
      \textbf{Label} & \textbf{Type} & \textbf{Condition} & \textbf{Short description} & \textbf{Link} \\
      \midrule
      \hypertarget{vid:S1}{S1} & Benchtop & Raw & Multi-view raw benchtop footage of the silicone vessel experiment before downstream reconstruction and metric analysis. & \href{https://drive.google.com/file/d/1ciIZe66ouoPVLSfL8lS-jLiJ6usg0wPZ/view}{Link} \\
      \hypertarget{vid:S2}{S2} & Benchtop & Cervical displacement & Displacement visualization for the benchtop cervical aspiration-catheter placement condition. & \href{https://drive.google.com/file/d/1Bn4HBy0Hwr1SYiJB4gT1hebgWc5V9Dzv/view}{Link} \\
      \hypertarget{vid:S3}{S3} & Benchtop & Terminal displacement & Displacement visualization for the benchtop ICA-terminus aspiration-catheter placement condition. & \href{https://drive.google.com/file/d/1hmIpD3wzWZ7jl9Tu8FZ1sh51d06xD7s6/view}{Link} \\
      \hypertarget{vid:S4}{S4} & Benchtop & Cervical stress proxy & Stress-proxy visualization for the benchtop cervical aspiration-catheter placement condition. & \href{https://drive.google.com/file/d/1tiha8fr2mQx9oL5wN31C1m8Ask7gf8WU/view}{Link} \\
      \hypertarget{vid:S5}{S5} & Benchtop & Terminal stress proxy & Stress-proxy visualization for the benchtop ICA-terminus aspiration-catheter placement condition. & \href{https://drive.google.com/file/d/1W4vKp3ptMvLMFJkkjRTaucRohOv8x6CP/view}{Link} \\
      \hypertarget{vid:S6}{S6} & Synthetic & Bulk raw & Raw rendered sequence for the synthetic bulk-translation control experiment. & \href{https://drive.google.com/file/d/1g7_ABaeUBfYnb4v1B5qirjQ4dnz2o1f5/view?usp=drive_link}{Link} \\
      \hypertarget{vid:S7}{S7} & Synthetic & Bulk displacement & Displacement visualization for the synthetic bulk-translation control experiment. & \href{https://drive.google.com/file/d/1QHWcEjLj_rrgQSqEt-NWeErXsbGIYf-t/view?usp=drive_link}{Link} \\
      \hypertarget{vid:S8}{S8} & Synthetic & Bulk stress proxy & Stress-proxy visualization for the synthetic bulk-translation control experiment. & \href{https://drive.google.com/file/d/1cTpkE3IXIMSyKZZe1YlaPgTB8LHFEkrj/view?usp=drive_link}{Link} \\
      \hypertarget{vid:S9}{S9} & Synthetic & Pulling raw & Raw rendered sequence for the synthetic localized-pulling validation experiment. & \href{https://drive.google.com/file/d/1eplMa8ViWlkR6__cXFebLH4Lmpq-0aQO/view?usp=drive_link}{Link} \\
      \hypertarget{vid:S10}{S10} & Synthetic & Pulling displacement & Displacement visualization for the synthetic localized-pulling validation experiment. & \href{https://drive.google.com/file/d/17ZqLaPUPnW0hhWYtM6piiLirpyLQxLV2/view?usp=drive_link}{Link} \\
      \hypertarget{vid:S11}{S11} & Synthetic & Pulling stress proxy & Stress-proxy visualization for the synthetic localized-pulling validation experiment. & \href{https://drive.google.com/file/d/1M54rr8L20nJyKrnCvgkVhYwt0otOeG7t/view?usp=drive_link}{Link} \\
      \bottomrule
    \end{tabular}%
  }
\end{minipage}

\fi

\end{document}